\pdfoutput=1
\documentclass[12pt,titlepage]{article}

\font\grande=cmr9.5 scaled \magstep4
\font\medio=cmr9.5 scaled \magstep2
\outer\def\beginsection#1\par{\medbreak\bigskip
      \message{#1}\leftline{\bf#1}\nobreak\medskip
\vskip-\parskip
      \noindent}
\usepackage{graphicx} 
\begin{document}
\bibliographystyle {unsrt}

\titlepage

\begin{flushright}
CERN-PH-TH-2019-131
\end{flushright}

\vspace{1cm}
\begin{center}
{\grande Effective energy density of relic gravitons}\\
\vspace{1cm}
Massimo Giovannini \footnote{e-mail address: massimo.giovannini@cern.ch}\\
\vspace{1cm}
{{\sl Department of Physics, CERN, 1211 Geneva 23, Switzerland }}\\
\vspace{0.5cm}
{{\sl INFN, Section of Milan-Bicocca, 20126 Milan, Italy}}
\vspace*{1cm}
\end{center}
\vskip 0.3cm
\centerline{\medio  Abstract}
\vskip 0.1cm
The energy density, the pressure and the anisotropic stress of the relic gravitons do not
have a unique gauge-invariant and frame-invariant expression since the equivalence principle 
ultimately forbids the localization of the energy-momentum of the gravitational field. 
All the strategies proposed so far suggest compatible descriptions inside the effective horizon 
but they lead to sharply different answers when the wavelengths of the gravitons exceed 
the Hubble radius by remaining shorter than the typical extension of causally connected patches. 
We present a coherent discussion of the energy-momentum pseudo-tensors of the relic gravitons in the 
different kinematical regimes with the aim of scrutinizing the mutual consistency of the competing 
suggestions. The various proposals are systematically compared by deducing and analyzing 
the explicit form of the observables in realistic physical situations. General lessons are drawn on the most 
plausible parametrization of the energy-momentum pseudo-tensor of the relic gravitons. 
\noindent
\vspace{5mm}
\vfill
\newpage
\renewcommand{\theequation}{1.\arabic{equation}}
\section{Introduction}
\setcounter{equation}{0}
\label{sec1}
The energy and the momentum of the gravitational field cannot be 
localized \cite{one,two}. In fact, assuming the equivalence principle in its 
stronger formulation, the laws of physics are those of special relativity in 
a freely falling (non-rotating) laboratory that occupies a small portion of spacetime. 
As long as the coordinates can be transformed to a freely falling frame,  
it is possible to eliminate locally the gravitational field. It is then conceptually 
difficult to propose a unique and local definition of the gravitational energy density.  
As a consequence there are no reasons why the energy-momentum 
tensor of the gravitational field itself should be either unique or 
covariantly conserved. For the same reason the energy-momentum tensor 
of the gravitational waves of cosmological origin is not unique. 
A variety of pseudo-tensors can be concocted and they should be
ultimately equivalent at least in the short wavelength limit, i.e. when the 
typical frequencies are much larger than the rate of variation of the corresponding geometry. 
This equivalence is not conclusive since there are physical situations where the frequencies 
of the waves are smaller than the rate of variation of the corresponding 
background. For instance, if we consider cosmological backgrounds during an accelerated 
stage of expansion the particle horizon diverges while the event horizon is proportional to the 
Hubble rate. The wavelengths  of the gravitons become  larger than the 
Hubble radius but are still shorter than the typical size of  causally connected 
regions:  by a mode being beyond the horizon we only mean 
that the physical wavenumber $k/a$ is much less than the expansion rate $H$ and
this does not have anything to do with causality \cite{threea}.

Some of the most notable strategies developed through the years can be 
summarized, in short, as follows. The Landau and Lifshitz \cite{three} proposal 
is rooted in the second-order corrections of the Einstein tensor supplemented 
by the observation that Bianchi identities must be valid to all orders in the perturbative expansion.
The Brill-Hartle strategy \cite{four} is instead based on the properties of a specific covariant averaging 
scheme aimed at separating the terms that evolve faster than the rate of 
variation of the corresponding background. The Brill-Hartle scheme has been used to derive
the Isaacson effective pseudo-tensor providing a sound description 
of gravitational radiation in the high-frequency limit  \cite{five} (see also \cite{fivea}).
The suggestion of Ford and Parker \cite{six,seven} follows instead from the  
effective action of the relic gravitons derived by perturbing the gravitational action 
to second order in the tensor amplitude. Other apparently different strategies are related to the 
ones mentioned above. For instance the approach of Refs. \cite{eight,nine} is 
the Landau-Lifshitz approach appropriately discussed in the case of a cosmological background. 

Through the years the various suggestions have been tested in different frameworks 
either for the solution of the concrete problems of backreaction \cite{nine,ten} or for 
the analysis of the implications of the different proposals \cite{eleven,twelve,thirteen,fourteen}
not necessarily in connection with the cosmological problems. 
Babak and Grishchuk \cite{fifteen} came up with a possible 
definition of a true energy-momentum tensor of the gravitational field. 
By treating gravity as a nonlinear tensor field in flat space-time \cite{sixteen}, 
 Ref. \cite{fifteen} claimed a result with all the 
necessary properties of a true energy-momentum tensor of the 
gravitational field itself (i.e. symmetry, uniqueness, gauge-invariance and covariant conservation). 
By taking the results of Ref.\cite{fifteen} at face value 
the problem of localizing the energy and momentum of the gravitational field would be 
completely solved.  The perspective of Ref. \cite{fifteen} has been subsequently questioned by Butcher, Hobson and Lasenby \cite{seventeen,eighteen,nineteen} who suggested that the proposal of Ref. \cite{fifteen} 
does not have a definite physical significance. In spite of the reasonable concerns of Refs. \cite{seventeen,eighteen,nineteen}, 
what matters for the present ends is that the geometrical object most closely related 
to the Babak-Grishchuk suggestion is (again) the Landau-Lifshitz pseudo-tensor \cite{three} as explicitly 
recognized by the authors of Ref. \cite{fifteen}. 

In this paper we intend to clarify the analogies and the important differences characterizing the 
various approaches developed so far. After scrutinizing the limitations and the ambiguities
of the diverse proposals some general lessons will be drawn. In the light of a number of reasonable 
criteria (i.e. gauge-invariance, frame-invariance, positivity of the energy density) a rather 
plausible strategy to assign the energy-momentum tensor of the relic gravitons is rooted in 
 the original Ford-Parker suggestion \cite{six,seven} where the background metric and the 
 corresponding perturbations are treated as independent fields;  the effective
 energy-momentum pseudo-tensor follows by functional derivation 
 of the effective action with respect to the background metric. In view of the general discussion 
 it is practical to separate three complementary 
 aspects of the problem: {\it i)} the strategy for the derivation of the energy-momentum pseudo-tensor; {\it ii)} 
 the averaging scheme; {\it iii)} the connection to the observables. 
 This will be the overall logic followed in the present investigation.

The pseudo-tensors explored through the years must implicitly assume an averaging scheme which is often difficult to formulate in general terms.
As long as the relic gravitons potentially populating the present universe did start their evolution either from 
a quantum mechanical initial state 
the expectation values of their energy density and of their pressure can be computed without 
imposing any extrinsic averaging scheme. Indeed during an inflationary stage of expansion the classical 
fluctuations are diluted away \cite{twentythreea,twentythreeb,twentythreec,twentythreed}
while the quantum fluctuations reappear continuously so that the relic gravitons are parametrically 
amplified thanks to the pumping action of the gravitational field itself, a perspective 
invoked in Refs.  \cite{twenty,twentyone} (see also \cite{twentythree}) 
even before the formulation of the conventional inflationary paradigm.  
By following the tenets of the quantum theory of parametric amplification 
(originally developed in the case of optical photons \cite{twentytwo}) a fair estimate of the mean energy density 
and pressure of the relic gravitons is obtained by averaging the various expressions over the same initial state 
(e.g. the vacuum).  Within each of the various parametrizations 
of the energy-momentum pseudo-tensor the quantum averages will then be used to compare the competing proposals. 
The other schemes (like the Brill-Hartle average  \cite{four} and its descendants \cite{eleven,twelve,fourteen}) 
reproduce the results of the quantum averaging when the wavelengths are shorter than the Hubble radius
but are not defined in the opposite limit.

The present paper is organized as follows. In section \ref{sec2} the various proposals are presented in a common 
perspective and with the purpose of easing their mutual comparison. By focussing the attention on the
case of cosmological backgrounds the explicit expressions for the energy density, for the pressure and for the anisotropic stress are obtained 
 within the different physical suggestions. The quantum averaging and its basic properties is discussed in section \ref{sec3}.
 Two  physically relevant examples are presented in section \ref{sec4} with the aim of illustrating 
 the basic features and the patent ambiguities of the different parametrizations.  
 The connection of the various proposals with the observables customarily 
employed in the analysis of relic graviton backgrounds is discussed in section \ref{sec5}. In the same framework we also
 evaluate the spectral density in the Brill-Hartle-Isaacson approach by assuming that the tensor amplitude is just an isotropic 
 random field varying in time; we connect the related spectral density to the power spectrum. At the end of section \ref{sec5} 
we show that the effective energy-momentum pseudo-tensor can be derived in a different (conformally related) frame
even if the obtained results are ultimately frame-invariant. Finally section \ref{sec6} contains the concluding remarks.

\renewcommand{\theequation}{2.\arabic{equation}}
\section{Different answers for a similar question}
\setcounter{equation}{0}
\label{sec2}
Let us consider a conformally 
flat background geometry $\overline{g}_{\mu\nu} = a^2(\tau) \eta_{\mu\nu}$, 
where $a(\tau)$ is the scale factor and $\tau$ denotes the conformal time coordinate;  
$\eta_{\mu\nu}= \mathrm{diag}(1, \, -1, \, -1,\, -1)$ is the Minkowski metric. 
The metric fluctuations are introduced as  $g_{\mu\nu}(\vec{x}, \tau) = \overline{g}_{\mu\nu} + \delta^{(1)}_{t} g_{\mu\nu}$, 
where  $\delta_{t}^{(1)} g_{\mu\nu}$ denotes
the (first-order) tensor fluctuation. The same notation will be used 
for any tensor in four-dimensions (i.e. the Ricci or Einstein tensor) so that 
$\delta_{t}^{(1)} A_{\mu\nu}$ and $\delta_{t}^{(2)}A_{\mu\nu}$ will denote the first- and second-order tensor fluctuations 
of the generic tensor $A_{\mu\nu}$. The first- and second-order tensor fluctuations of the metric and of the square root of its determinant 
are given by:
\begin{eqnarray}
\delta_{t}^{(1)}\, g_{ij} &=& - a^2\, h_{ij}, \qquad \delta_{t}^{(1)} \, g^{ij} = \frac{h^{ij}}{a^2}, \qquad 
\delta_{t}^{(2)} \, g^{ij} = - \frac{h^{i}_{k}\,\, h^{j k}}{a^2},
\label{oneB}\\
\delta_{t}^{(1)} \sqrt{ -g} &=&0, \qquad \delta_{t}^{(2)} \sqrt{ -g} = - \frac{a^4}{4} h_{k\ell}\, h^{k\ell},
\label{oneBB}
\end{eqnarray}
where $h_{i\,j}$ is a rank-two tensor in three dimensions which is divergenceless and traceless, i.e. 
$\partial_{i} h^{\,\,i}_{j} =0 = h_{i}^{\,\,\,i}$. The prime will denote a derivation with respect to the 
conformal time coordinate. With this notation ${\mathcal H} = a^{\prime}/a$ where ${\mathcal H} = a \, H$ and $H = \dot{a}/a$ 
is the Hubble rate in the cosmic time parametrization [note that $a(\tau) \, d\tau = d\, t$]. 
To avoid the proliferation of superscripts we shall sometimes make explicit the derivation with respect to $\tau$ 
and write $h_{k\ell} \,\partial_{\tau} h^{k\ell}$ instead of $h_{k\ell} \,h^{k\ell\,\,\prime}$. It is relevant to mention 
that the tensor amplitude $h_{ij}$ defined in Eqs. (\ref{oneB})-(\ref{oneBB}) is invariant under 
infinitesimal coordinate transformations; if the tensor amplitude is defined as in Eqs. (\ref{oneB})--(\ref{oneBB})
its quadratic combinations will be automatically gauge-invariant. 

\subsection{The effective action of the relic gravitons}
The effective energy-momentum pseudo-tensor of the relic gravitons follows from the observation 
\cite{six,seven} that the tensor fluctuations and the background 
metric can be regarded as independent variables.  Neglecting, for simplicity, the presence of the sources the 
action for the relic gravitons is essentially the Einstein-Hilbert action perturbed to second-order i.e.
\begin{equation}
S_{t} = \delta^{2} S = \frac{1}{2 \ell_{P}^2} \delta^{(2)}_{t}  \biggl\{\int d^{4} x \biggl[ \sqrt{- g} \, g^{\alpha\beta} \,\biggl(\Gamma_{\alpha\beta}^{\,\,\,\,\, \rho} \,\,
\Gamma_{\rho\sigma}^{\,\,\,\,\, \sigma} - \Gamma_{\beta\rho}^{\,\,\,\,\, \sigma} \,\,
\Gamma_{\sigma\alpha}^{\,\,\,\,\, \rho} \biggr) \biggr]\biggr\},
\label{twoA}
\end{equation}
where $\ell_{P} = \sqrt{8 \pi G}$ and the quantity appearing inside the curly bracket is the Einstein-Hilbert action where 
the total derivatives have been excluded. The second-order fluctuation implicitly indicated in Eq. (\ref{twoA}) can also be expressed as
\begin{eqnarray}
 \delta^{(2)}_{t} S &=& \frac{1}{2 \ell_{P}^2}\int d^{4} x  \biggl[  \overline{g}^{\alpha\beta} \, \overline{{\mathcal Z}}_{\alpha\beta} \,\,\delta^{(2)}_{t} \sqrt{-g} + \sqrt{ -\overline{g}} \biggl( \delta^{(2)}_{t} g^{\alpha\beta} \,\overline{{\mathcal Z}}_{\alpha\beta} 
 \nonumber\\
 &+&
 \delta^{(1)}_{t} g^{\alpha\beta} \,  \delta^{(1)}_{t} {\mathcal Z}_{\alpha\beta} + \overline{g}^{\alpha\beta}\,\,\delta^{(2)}_{t} {\mathcal Z}_{\alpha\beta} \biggr)\biggr],
 \label{twoB}
 \end{eqnarray}
where ${\mathcal Z}_{\alpha\beta} = \Gamma_{\alpha\beta}^{\,\,\,\,\, \rho} 
\Gamma_{\rho\sigma}^{\,\,\,\,\, \sigma} - \Gamma_{\beta\rho}^{\,\,\,\,\, \sigma} \Gamma_{\sigma\alpha}^{\,\,\,\,\, \rho}$
and $\overline{{\mathcal Z}}_{\alpha\beta}$ denotes the corresponding background quantity. To zeroth-order we have that  $\overline{{\mathcal Z}}_{00} =0$ and $\overline{{\mathcal Z}}_{ij} = 2 {\mathcal H}^2 \delta_{ij}$. To first-order 
in the tensor amplitude we have instead $\delta^{(1)} {\mathcal Z}_{00} =0$ while $\delta^{(1)} {\mathcal Z}_{ij} = 2 {\mathcal H}^2 \, h_{ij}$. Finally the explicit second-order contributions are:
\begin{eqnarray}
\delta^{(2)} {\mathcal Z}_{00} &=& - \frac{1}{4} h_{k\ell}^{\,\prime} \,  h^{k\ell\,\,\prime} + \frac{{\mathcal H}}{2} h_{k\ell}^{\prime}\, h^{k\ell},
\label{twoBa}\\
\delta^{(2)} {\mathcal Z}_{ij} &=& - \frac{{\mathcal H}}{2} h_{k\ell}^{\prime}\, h^{k\ell} \,\,\delta_{ij} - \frac{1}{4} \biggl[ h_{i}^{\,\,k\,\,\prime}\,\, h_{k\, j}^{\prime} + h_{j}^{\,\,k\,\,\prime}\,\, h_{k\, i}^{\prime}\biggr] 
\nonumber\\
&-& \frac{1}{4} \biggl[ \partial_{\ell} \,h_{i}^{\,\, k} + \partial_{i} h_{\ell}^{\,\, k} - \partial^{k} \,h_{\ell \,i} \biggr] \biggl[ \partial_{k} \,\,h_{j}^{\,\, \ell} + \partial_{j} \,\,h_{k}^{\,\, \ell} - \partial^{\ell} \,\,h_{k \,j} \biggr].
\label{twoBb}
\end{eqnarray}
Inserting Eqs. (\ref{twoBa}) and (\ref{twoBb}) into Eq. (\ref{twoB}) the effective action of the relic gravitons is:
\begin{equation}
S_{t} = \frac{1}{8 \ell_{P}^2} \int d^{4} x \sqrt{ - \overline{g}} \,\, \overline{g}^{\alpha\beta} \, \, \partial_{\alpha} h_{ij} \, \partial_{\beta} h^{ij}.  
\label{threeB}
\end{equation}
The possible presence of background sources does not change the result of Eq. (\ref{threeB}).
In fact $\delta^{(2)}_{t} S$ must always be evaluated by imposing the validity of the background evolution 
and the tensor modes decouple from the matter fields at least if the anisotropic stress of the sources 
vanishes. Since the effective action of the relic gravitons in a conformally flat metric is given by Eq. (\ref{threeB}),
their energy-momentum pseudo-tensor can be introduced from the functional derivative 
of $S_{t}$ with respect to $\overline{g}_{\mu\nu}$ by considering $h_{ij}$ and $\overline{g}_{\mu\nu}$ as 
independent variables:
\begin{equation}
\delta S_{t} = \frac{1}{2} \int d^{4} x\,\, \sqrt{-\overline{g}} \,\, T^{(gw)}_{\mu\nu}  \,\,\delta \overline{g}^{\mu\nu}.
\label{fourB}
\end{equation} 
From Eq. (\ref{threeB}) the explicit form of Eq.  (\ref{fourB}) becomes: 
 \begin{equation}
{\mathcal F}_{\mu\nu} = \frac{1}{4 \ell_{\mathrm{P}}^2} \biggl[ \partial_{\mu} h_{i j} \,\,\partial_{\nu} h^{i j} 
- \frac{1}{2} \overline{g}_{\mu \nu} \,\,\biggl(\overline{g}^{\alpha\beta}\, \partial_{\alpha} h_{ij} \,\,\partial_{\beta} h^{ij} \biggr)\biggr],
\label{fiveB}
\end{equation}
where we used the notation ${\mathcal F}_{\mu\nu} = T_{\mu\nu}^{(gw)}$ to distinguish Eq. (\ref{fiveB}) from the other proposals examined below. The indices of ${\mathcal F}_{\mu\nu}$ 
are raised and lowered with the help of the background metric (i.e. ${\mathcal F}_{\mu}^{\nu} = \overline{g}^{\alpha\nu} {\mathcal F}_{\alpha\nu}$); the energy density and the pressure are:
\begin{eqnarray}
\rho^{(F)}_{gw} &=& \frac{1}{8 \ell_{\mathrm{P}}^2 a^2} \biggl[ \partial_{\tau} h_{k \ell}\, \partial_{\tau}h^{k \ell} + \partial_{m} h_{k\ell} \partial^{m} h^{k\ell}\biggr],
\label{rhoF}\\
p^{(F)}_{gw} &=&   \frac{1}{8 \ell_{\mathrm{P}}^2 a^2} \biggl[ \partial_{\tau} h_{k\ell}\partial_{\tau} h^{\,k\ell} - \frac{1}{3} \partial_{m} h_{k \ell} \,\partial^{m} h^{\,k\ell}  \biggr].
\label{pF}
\end{eqnarray}
The associated anisotropic stress is traceless (i.e. $\Pi_{i}^{(F)\,\,i} =0$) and it is:
\begin{equation}
\Pi_{i}^{(F)\,\,j} = \frac{1}{4 \ell_{\mathrm{P}}^2 a^2} \biggl[ - \partial_{i} \,h_{k\ell} \,\,\partial^{j} \,h^{k\ell} + \frac{1}{3} \delta_{i}^{j} \,\,\partial_{m}\, h_{k \ell} \,\,\partial_{m} \,h_{k\ell} \biggr].
\label{anF}
\end{equation}
In terms of Eqs. (\ref{rhoF}), (\ref{pF}) and (\ref{anF}) the components of ${\mathcal F}_{\mu}^{\nu}$ are 
\begin{eqnarray}
{\mathcal F}_{0}^{0} &=& \rho^{(F)}_{gw}, \qquad {\mathcal F}_{i}^{0} = S_{i}^{(F)}= 
\frac{1}{4 \ell_{\mathrm{P}}^2 a^2} \partial_{\tau} h_{k\ell} \,\partial_{i} h^{k\ell},
\nonumber\\
{\mathcal F}_{i}^{j} &=& - p_{gw}^{(F)} \,\,\delta_{i}^{j} + \Pi_{i}^{(F)\,\,j},
\label{sixB}
\end{eqnarray}
where $ S_{i}^{(F)}$ denotes the energy flux. The energy density, the pressure and the energy flux
combine in the following identity 
\begin{eqnarray}
 \partial_{\tau} \rho^{(F)}_{gw} + 3 {\mathcal H} \biggl[\rho^{(F)}_{gw}  + p^{(F)}_{gw} \biggr] = \frac{h_{k\ell}^{\prime}}{4 \ell_{P}^2 a^2}
[ h_{k \ell}^{\prime\prime} + 2 {\mathcal H} h_{k\ell}^{\prime} - \nabla^2 h_{k \ell} ] + \vec{\nabla} \cdot \vec{S}^{(F)}.
\label{sevenB}
\end{eqnarray}
The first term at the right hand side of Eq. (\ref{sevenB}) vanishes because of the evolution of the 
tensor amplitude following from the extremization of the action (\ref{threeB}) with respect to $h_{ij}$.  
The second term at the right hand side of Eq. (\ref{sevenB}) is not vanishing, in general; but if we regard the energy flux as an operator constructed from the corresponding quantum fields, its expectation value over the vacuum is generally vanishing (see section \ref{sec3}).

\subsection{The second-order variation of the Einstein tensor}
The Landau-Lifshitz strategy that is based on the 
analysis of the nonlinear corrections to the Einstein tensor consisting, to lowest order,  
of quadratic combinations of the tensor amplitude $h_{ij}$. While the derivation of Eq. (\ref{fiveB}) does not require the systematic use of 
a the evolution of the tensor amplitude the opposite is true in the Landau-Lifshitz framework 
where the energy-momentum pseudo-tensor ${\mathcal L}_{\mu}^{\nu}$  can be expressed as: 
\begin{equation}
\ell_{\rm P}^2 {\mathcal L}_{\mu}^{\,\,\,\nu} = - \delta^{(2)}_{t} {\mathcal G}_{\mu}^{\,\,\,\nu}.
\label{oneC}
\end{equation}
Furthermore, since the Bianchi identity $\nabla_{\mu} {\cal G}_{\nu}^{\mu}=0$
must be valid to all orders, we must also demand that $\delta_{\rm t}^{(2)} ( \nabla_{\mu} {\cal G}^{\mu}_{\nu}) =0$
ultimately leading to a relation analog to Eq. (\ref{sevenB}). From the second-order fluctuations 
 of the Einstein tensor the energy density, the pressure and the anisotropic stress are:
 \begin{eqnarray}
\rho_{gw}^{(L)} &=&  \frac{1}{a^2 \ell_{\rm P}^2} \biggl[ {\mathcal H} \,
(\partial_{\tau}h_{k\ell })\, h^{k\ell} + \frac{1}{8} ( \partial_{m} h_{k\ell}\,\,\partial^{m} h^{k\ell} + 
\partial_{\tau} h_{k\ell}\,\, \partial_{\tau} h^{k\ell})\biggr],
\label{rhoL}\\
p_{gw}^{(L)} &=& - \frac{1}{24 a^2 \ell_{\rm P}^2}\biggl[ 5 \,\partial_{\tau}h_{k\ell}\,\partial_{\tau}h^{k\ell} - 7\,\,
\partial_{m} h_{k\ell}\, \partial^{m} h^{k\ell} \biggr].
\label{pL}\\
\Pi_{i}^{(L)\,\,j} &=& 
\frac{1}{a^2 \ell_{P}^2} \biggl\{ \frac{1}{6} \biggl[ \partial_{\tau}\,h_{k\ell}\, \partial_{\tau}\,h^{k\ell} - 
\frac{1}{2} \partial_{m}\, h_{k\ell} \,\,\partial^{m}\, h^{k\ell} \biggr] \delta_{i}^{j}
+ \frac{1}{2} \partial_{m} \,h_{\ell i} \,\,\partial^{m}\, h^{\ell j} 
\nonumber\\
&-& \frac{1}{4} \partial_{i}\, h_{k\ell} \,\,\partial^{j}  \,h^{k\ell} 
- \frac{1}{2} \partial_{\tau}\,h_{k i}\,\, \partial_{\tau}\, h^{k j} \biggr\},
\label{fourC}
\end{eqnarray}
 with $\Pi_{i}^{(L)\,\,i} =0$. In analogy with Eq. (\ref{sixB}) the components of energy-momentum pseudo-tensor ${\mathcal L}_{\mu}^{\nu}$
 in the Landau-Lifshitz approach are:
\begin{eqnarray}
{\mathcal L}_{0}^{0} &=& \rho^{(L)}_{gw}, \qquad  {\mathcal L}_{i}^{0} = S_{i}^{(L)}  =
\frac{1}{4 \ell_{\mathrm{P}}^2 a^2} \partial_{\tau} \,h_{k\ell} \,\,\partial_{i}\, h^{k\ell},
\nonumber\\
{\mathcal L}_{i}^{j} &=& - p_{gw}^{(L)} \delta_{i}^{j} + \Pi_{i}^{(L)\,\,j}.
\label{fiveC}
\end{eqnarray}
For a more direct comparison with Eqs. (\ref{rhoF}) and (\ref{pF}) 
 various total derivatives (i.e. three-divergences of a quadratic combination of tensor amplitudes) have been excluded 
 from Eqs. (\ref{rhoL}) and (\ref{pL}). Consider, for instance, the second-order variations of the Ricci tensor and of the 
Ricci scalar contributing to $\delta_{t}^{(2)} {\mathcal G}_{00}$ (and to ${\mathcal L}_{00}$):
\begin{eqnarray}
\delta_{t}^{(2)} R_{00} &=& \frac{1}{4} \partial_{\tau} h_{k\ell} \,\,\partial_{\tau} h^{k \ell} + \frac{1}{2} h^{k \ell} \,\,[ h_{k \ell}^{\prime\prime} + {\mathcal H} h_{k \ell}^{\prime} ],
\label{sixC}\\
\delta_{t}^{(2)} R &=& \frac{1}{a^2}\biggl[ \frac{3}{4} \partial_{\tau} h_{k\ell}  \partial_{\tau} {h^{k\ell}} + 
{\mathcal H} \partial_{\tau} h_{k\ell} h^{k\ell} - \frac{3}{4} \partial_{i}h^{k\ell} \partial^{i} h_{k\ell}\biggr]
+ \frac{1}{a^2} {\mathcal D}_{R}, 
\nonumber\\
{\mathcal D}_{R} &=& \partial_{i} \biggl[ h_{k \ell} \partial^{i} h^{k \ell} - \frac{1}{4} h_{k}^{ \ell} \partial_{\ell} h^{k}_{i} \biggr],
\label{sevenC}
\end{eqnarray}
where ${\mathcal D}_{R}$ is the total derivative term. According to the logic of this approach
the term $h_{k \ell}^{\,\prime\prime}$ must be replaced by $ - 2 {\mathcal H} h_{k \ell}^{\prime} + \nabla^2 h_{k \ell}$
that follows from the evolution of the tensor amplitude.  As a result of this lengthy but straightforward procedure
when Eqs. (\ref{sixC}) and (\ref{sevenC}) are combined in 
${\mathcal L}_{00}$ a further total derivative term emerges so that the final result for the energy density is:
\begin{equation}
{\mathcal L}_{00} = \frac{1}{\ell_{P}^2} \biggl[ ({\mathcal H} \,
\partial_{\tau}h_{k\ell })\, h^{k\ell} + \frac{1}{8} ( \partial_{m} h_{k\ell}\,\partial^{m} h^{k\ell} + 
\partial_{\tau} h_{k\ell} \,\partial_{\tau} h^{k\ell})\biggr] - \frac{1}{8 \ell_{P}^2} {\mathcal D}_{00},
\label{eightC}
\end{equation}
where $ {\mathcal D}_{00} = \partial_{i} [ h_{k \ell}\, \partial^{\ell} h^{k \, i} ]$. 
All in all Eq. (\ref{eightC}) shows, as anticipated, that Eq. (\ref{rhoL}) is determined up to the 
total derivative term (i.e. ${\mathcal D}_{00}$) and the same  happens in the case of 
 the pressure terms whose associated total derivatives are qualitatively similar to ${\mathcal D}_{R}$ and ${\mathcal D}_{00}$ 
 but will not be explicitly reported.  Finally the explicit form of the condition $\delta_{t}^{(2)} ( \nabla_{\mu} {\cal G}^{\mu}_{\nu}) =0$  
 following from the validity of the Bianchi identity is:
\begin{eqnarray}
&&\partial_{\mu} \delta_{t}^{(2)} {\mathcal G}^{\,\,\mu}_{\nu} + 
\delta_{t}^{(2)} \Gamma_{\mu\alpha}^{\,\,\,\,\,\mu} \overline{\mathcal G}_{\,\,\nu}^{\alpha} + 
\overline{\Gamma}_{\mu\alpha}^{\,\,\,\,\,\,\mu} \delta^{(2)}_{t} {\mathcal G}_{\,\,\nu}^{\alpha} +
\delta_{ t}^{(1)} \Gamma_{\mu\alpha}^{\,\,\,\,\,\mu} \delta^{(1)}_{t} {\mathcal G}_{\nu}^{\,\,\alpha} 
- \delta_{t}^{(2)} \Gamma_{\nu\alpha}^{\,\,\,\,\,\beta} \overline{{\mathcal G}}_{\beta}^{\,\,\alpha}
\nonumber\\
&& - \overline{\Gamma}_{\nu\alpha}^{\,\,\,\,\,\,\,\beta} \delta_{t}^{(2)} {\mathcal G}_{\beta}^{\,\,\alpha} -
\delta_{ t}^{(1)} \Gamma_{\nu\alpha}^{\,\,\,\,\,\beta} \delta^{(1)}_{ t} {\mathcal G}_{\beta}^{\,\,\alpha}=0.
\label{nineC}
\end{eqnarray}
Equation (\ref{nineC}) implies some sort of conservation equation similar to Eq. (\ref{sevenB}); indeed, from 
the energy density and the pressure defined in Eqs. (\ref{rhoL}) and (\ref{pL}), f Eq. (\ref{nineC}) becomes
after some algebra:
\begin{equation}
\partial_{\tau} \rho_{gw}^{(L)} + 3 {\mathcal H}\biggl[ \rho_{gw}^{(L)} + {\mathcal P}_{gw}^{(L)}\biggr] 
= \frac{1}{ 4 \ell_{P}^2 a^2} [h_{k \ell}^{\prime} + 4 {\mathcal H} h_{k \ell} ][ h_{k\ell}^{\prime\prime} 
+ 2 {\mathcal H} h_{k \ell}^{\prime} - \nabla^2 h_{k\ell} ] + \vec{\nabla} \cdot \vec{Q}^{(L)}.
\label{tenC}
\end{equation}
In Eq. (\ref{tenC})  the shifted pressure ${\mathcal P}_{gw}^{(L)}$  does not 
coincide with $p_{gw}^{(L)}$ (see Eq. (\ref{pL})); the same comment holds for $\vec{Q}^{(L)}$ which differs from $\vec{S}^{(L)}$ introduced in Eq. (\ref{fiveC}). In explicit terms we have that the shifted pressure and the shifted vector are:
\begin{eqnarray}
 {\mathcal P}_{gw}^{(L)} &=& p_{gw}^{(L)} +  \frac{({\mathcal H}^2 - {\mathcal H}^{\prime})}{ 3 {\mathcal H} a^2\ell_{P}^2} (\partial_{\tau} h_{k\ell}) h^{k\ell},
\label{elevenC}\\
Q_{i}^{(L)} &=& \frac{1}{4 \ell_{P}^2 a^2} [h_{k\ell}^{\prime} + 4 {\mathcal H} h_{k \ell}] \,\,\partial_{i}  h^{k \ell}.
\label{twelveC}
\end{eqnarray}
In comparison with $p_{gw}^{(L)}$ the value of ${\mathcal P}_{gw}^{(L)}$  is shifted 
 by the second-order fluctuations of the Christoffel connection 
\begin{equation}
{\mathcal P}_{gw}^{(L)} - p_{gw}^{(L)} = - \frac{2}{3 \,a^2 \, {\mathcal H}\, \ell_{P}^2} ({\mathcal H}^2 - {\mathcal H}^{\prime}) \delta_{t}^{(2)} \Gamma_{k0}^{\,\,\,\,\,k}, \qquad \delta_{t}^{(2)} \Gamma_{k0}^{\,\,\,\,\,k} = - \frac{1}{2} h_{k\ell} \partial_{\tau} h^{k \ell}.
\label{thirteenC} 
\end{equation}
The shifted pressure entering Eq. (\ref{tenC}) should be regarded as the true physical pressure
as it will emerge from the explicit examples of sections \ref{sec4} and \ref{sec5}.
The first term at the right hand side of Eq. (\ref{tenC}) vanishes because of the evolution of the first-order amplitude. 
The second term at the right-hand side of Eq. (\ref{tenC})  vanishes when averaged 
over the quantum state of the relic gravitons (see the discussion in section \ref{sec3}).

\subsection{The covariant approach}
The covariant approach, in its original formulation, assumes the Brill-Hartle scheme \cite{four} that implicitly 
selects the frequencies exceeding the Hubble rate. In the covariant approach the metric is decomposed as 
\begin{equation}
g_{\mu\nu} = \overline{g}_{\mu\nu} + \widetilde{h}_{\mu\nu}, \qquad u^{\mu} \widetilde{h}_{\mu\nu} =0, \qquad \overline{\nabla}_{\mu} \, \widetilde{h}^{\mu\nu} =0,\qquad \widetilde{h}_{\mu}^{\,\,\,\,\mu} =0,
\label{oneD}
\end{equation}
where $\overline{\nabla}_{\mu}$ denotes the covariant derivative with respect to the background metric 
$\overline{g}_{\mu\nu}$; the indices of $\widetilde{h}_{\mu\nu}$ are raised and lowered 
with the help of $\overline{g}_{\mu\nu}$. Within the approach of Eq. (\ref{oneD}) the 
cosmological fluctuations correspond to 
$u^{\mu} \widetilde{h}_{\mu\nu} =0$ where $u_{\mu}$ is the fluid four-velocity. In the case of a conformally flat background 
geometry $\overline{g}_{\mu\nu} = a^2(\tau) \eta_{\mu\nu}$ the conditions $u^{\mu} \widetilde{h}_{\mu\nu} =0$ and 
$ \overline{\nabla}_{\mu}\,\widetilde{h}^{\mu\nu} =0 $ imply $\widetilde{h}_{\mu} ^{\mu} =0$; if $a(\tau)$ is constant the three conditions 
must all be separately imposed\footnote{By projecting the condition $\overline{\nabla}_{\mu} \widetilde{h}^{\,\mu\nu}=0$ along $u_{\nu}$ we obtain, for a cosmological background with flat spatial sections, that  $(\overline{\nabla}_{\mu} \widetilde{h}^{\,\mu\nu})u_{\nu}= H (g_{\alpha\beta} - u_{\alpha} u_{\beta}) \widetilde{h}^{\alpha\beta}$ where $H$ is the Hubble rate. If we then impose, according to Eq. (\ref{oneD}) that $u^{\mu} \widetilde{h}_{\mu\nu} =0$, the condition 
 $(\overline{\nabla}_{\mu} \widetilde{h}^{\,\mu\nu}) u_{\nu} =0$ also demands $\widetilde{h}_{\mu}^{\,\,\mu} =0$ {\em provided} $H\neq 0$ [i.e. $a(\tau)$ must {\em not} be constant].}. The tensor amplitude $\widetilde{h}_{\mu\nu}$ of Eq. (\ref{oneD}) is related to the tensor amplitude $h_{ij}$ 
of Eqs. (\ref{oneB})--(\ref{oneBB}) as
\begin{equation}
\widetilde{h}_{ij} = - a^2 h_{ij}, \qquad \widetilde{h}_{0\mu} =0, \qquad \partial_{i} \widetilde{h}^{ij} =0, \qquad \widetilde{h}_{i}^{\,\,i} =0.
\label{twoD}
\end{equation}
Equation (\ref{twoD}) implies the conditions of Eq. (\ref{oneD}) but while the indices of $\widetilde{h}_{ij}$ are raised 
and lowered with the help of the background metric, the indices of $h_{ij}$ are all Euclidean. 
Within the covariant approach the energy-momentum tensor following from the Brill-Hartle
average is given by: 
\begin{equation}
{\mathcal B}_{\mu\nu} = \frac{1}{4 \ell_{P}^2} \overline{\nabla}_{\mu} \widetilde{h}_{\alpha\beta} \, \overline{\nabla}_{\nu} \, 
\widetilde{h}^{\alpha\beta}.
\label{threeD}
\end{equation}
To compare the covariant approach with the other proposals we insert Eq. (\ref{twoD}) 
inside Eq. (\ref{threeD}) and the result for the various components of ${\mathcal B}_{\mu}^{\,\,\,\nu}$ is:
\begin{eqnarray}
{\mathcal B}_{0}^{\,\,0} &=& \rho_{gw}^{(B)}, \qquad {\mathcal B}_{i}^{\,\,0} = S_{i}^{(B)} = \frac{1}{4 \ell_{P}^2 a^2} \, h_{k\ell}^{\prime}\,\,\partial_{i} \,h^{k \ell},
 \nonumber\\
{\mathcal B}_{i}^{\,\,j} &=& - p_{gw}^{(B)} \delta_{i}^{\,\, j} + \Pi_{i}^{(B)\,\, j}, 
\label{fourD}
\end{eqnarray}
where the energy density, the pressure and the anisotropic stress are:
\begin{eqnarray}
\rho_{gw}^{(B)} &=& \frac{1}{4 \ell_{P}^2 a^2} \partial_{\tau} h_{k \ell} \, \partial_{\tau} h^{k \ell},
\label{rhoB}\\
p_{gw}^{(B)} &=& \frac{1}{12 \ell_{P}^2 a^2} \partial_{m} h_{k \ell} \, \partial^{m} h^{k \ell},
\label{pB}\\
\Pi_{i}^{(B)\,\,j} &=&  \frac{1}{4 \ell_{P}^2 a^2} \biggl[ - \partial_{i} \,h_{k \ell} \,\,\partial^{j} \,h^{k \ell}  
+ \frac{1}{3} \delta_{i}^{\,\, j}\, \partial_{m} \, h_{k \ell}\,\, \partial^{m} \,h^{k \ell} \biggr].
\label{anB}
\end{eqnarray}
 Equation (\ref{threeD}) is the result of an averaging procedure which excludes by construction the long wavelengths and can therefore be applied only inside the Hubble radius. 
 If Eq. (\ref{threeD}) is blindly applied 
beyond the Hubble radius various ambiguities arise and they will be discussed in sections \ref{sec4} and \ref{sec5}.
The covariant approach can be however extended for typical wavelengths larger than the Hubble radius 
by employing a different averaging scheme. In this case the results applicable to cosmological background geometries 
will coincide exactly with Eq. (\ref{fiveB}) (see the discussion at the end of section \ref{sec3}).

\subsection{Mutual relations between the different prescriptions}
The expressions of the energy density obtained in the cases examined above do not coincide in general terms. To appreciate 
the  differences it is useful to write their mutual relations:
\begin{eqnarray}
\rho_{gw}^{(F)} &=& \rho_{gw}^{(L)} - \frac{{\mathcal H} h_{k\ell}^{\prime} \, h^{k \ell}}{a^2 \ell_{P}^2},
\label{FtoL}\\
\rho_{gw}^{(L)} &=& \frac{\rho_{gw}^{(B)}}{2} + \frac{{\mathcal H} \, h_{k\ell}^{\prime} \, h^{k \ell}}{a^2 \ell_{P}^2} + \frac{1}{8 \ell_{P}^2 a^2} \partial_{m} \,h_{k\ell} \,\,\partial^{m} \,h^{k\ell},
\label{LtoB}\\
\rho_{gw}^{(B)} &=& 2 \biggl[\rho_{gw}^{(F)} - \frac{1}{8 \ell_{P}^2 a^2} \partial_{m} \,h_{k\ell} \,\,\partial^{m} \,h^{k\ell}\biggr].
\label{BtoF}
\end{eqnarray}
The expressions of $\rho_{gw}^{(F)}$ and 
$\rho_{gw}^{(L)}$ are very similar but they differ by a crucial term containing ${\mathcal H}$. 
A similar remark holds in the case of the relation between $\rho_{gw}^{(B)}$ and $\rho_{gw}^{(F)}$ since they both
differ by terms that are negligible beyond the Hubble radius i.e. when the frequency of the graviton is smaller than the rate 
of variation of the background. The corresponding pressures obey qualitatively similar 
relations that can be easily deduced from the results given above. 

\renewcommand{\theequation}{3.\arabic{equation}}
\section{The quantum averaging}
\setcounter{equation}{0}
\label{sec3}
The classical and quantum fluctuations of cosmological backgrounds obey the  same evolution equations, 
but  while classical fluctuations are given once forever (on a given space-like hypersurface) quantum fluctuations 
keep on reappearing all the time. If the kinematical and dynamical problems of a decelerated cosmology are fixed 
by means of a phase of accelerated expansion lasting (at least) $65$ efolds, the classical fluctuation 
are exponentially suppressed during inflation \cite{twentythreea,twentythreeb,twentythreec,twentythreed} 
(see also \cite{twentythreee,twentythreef,twentythreeg}). At a purely classical level it is then plausible to conclude that
any finite portion of the event horizon gradually loses the memory of an initially imposed anisotropy 
or inhomogeneity so that the metric attains the observed regularity regardless of the initial boundary conditions.
Since in this situation the power spectra of the scalar and tensor modes of the geometry follow 
from the quantum mechanical expectation values of two field operators evaluated 
at the same time (but at different spatial locations), it is also very reasonable to 
apply the quantum averaging for the estimates of the expectation values of the
 different components of the pseudo-tensors derived above. 
 In the quantum description the classical fields and their 
derivatives are promoted to the status of quantum mechanical operators i.e. 
$h_{ij} \to \hat{h}_{ij}$ and $ h_{ij}^{\,\prime} \to \hat{h}_{ij}^{\, \prime}$:
\begin{eqnarray}
\hat{h}_{ij}(\vec{x}, \tau) &=& \frac{\sqrt{2} \, \ell_{P}}{(2\pi)^{3/2}} \sum_{\lambda} \int d^{3} k \,\, e^{(\lambda)}_{i\, j}(\hat{k})\biggl[ F_{k\,\lambda} \hat{a}_{\vec{k},\, \lambda} e^{- i \vec{k} \cdot\vec{x}}+ F_{k\,\lambda}^{*} \hat{a}_{\vec{k},\, \lambda}^{\dagger} e^{i \vec{k} \cdot\vec{x}} \biggr],
\label{oneE}\\
\hat{h}_{ij}^{\,\,\prime}(\vec{x}, \tau) &=& \frac{\sqrt{2} \, \ell_{P}}{(2\pi)^{3/2}} \sum_{\lambda} \int d^{3} k \,\, e^{(\lambda)}_{i\, j}(\hat{k})\biggl[ G_{k\,\lambda} \hat{a}_{\vec{k},\, \lambda} e^{- i \vec{k} \cdot\vec{x}}+ G_{k\,\lambda}^{*} \hat{a}_{\vec{k},\, \lambda}^{\dagger} e^{ i \vec{k} \cdot\vec{x}} \biggr],
\label{twoE}
\end{eqnarray}
where $[\hat{a}_{\vec{k},\, \lambda}, \, \hat{a}^{\dagger}_{\vec{p},\, \lambda^{\prime}}] = \delta_{\lambda\, \lambda^{\prime}} \delta^{(3)}
(\vec{k} - \vec{p})$; 
$e_{ij}^{(\lambda)}(\hat{k})$ (with $\lambda = \oplus, \, \otimes$) accounts for the two tensor polarizations\footnote{If we define a triplet of mutually orthogonal unit vectors 
$\hat{m}$, $\hat{n}$ and $\hat{k}$ we can set the direction 
of propagation of the wave along $\hat{k}$ and, in this case, 
the two tensor polarizations are 
$e_{ij}^{\oplus}= (\hat{m}_{i} \, \hat{n}_{j} - \hat{n}_{i} \, \hat{m}_{j})$ and 
$e_{ij}^{\otimes}= (\hat{m}_{i} \, \hat{n}_{j} + \hat{n}_{i} \, \hat{m}_{j})$.} and 
the mode functions (for each separate polarization) obey 
\begin{eqnarray}
&& G_{k} = F_{k}^{\,\,\prime}, \qquad G_{k}^{\,\,\prime} = - k^2 F_{k} - 2 {\mathcal H} G_{k},
\nonumber\\
&& F_{k}(\tau) G_{k}^{*}(\tau) - F_{k}^{*}(\tau) G_{k}(\tau) = \frac{i}{a^2(\tau)},
\label{threeE}
\end{eqnarray}
where $ F_{k, \oplus} = F_{k,\otimes} = F_{k}$ and $G_{k, \oplus} = G_{k,\otimes} = G_{k}$ in the unpolarized case treated here.
The sum over the polarizations is given by 
$\sum_{\lambda} e^{\lambda}_{i\, j} \, e^{(\lambda)}_{m\,n}(\hat{k}) = 4 {\mathcal S}_{i\,j\,m\, n}(\hat{k})$ 
and ${\mathcal S}_{i\, j\, m\, n}$ is defined as
\begin{equation}
{\mathcal S}_{i\,j\,m\, n} = \frac{1}{4} \biggl[ p_{i\,m}(\hat{k}) p_{j\, n}(\hat{k}) + p_{i\, n}(\hat{k}) p_{j\, m}(\hat{k}) - 
p_{i\,j}(\hat{k}) p_{m\, n}(\hat{k}) \biggr],
\label{Sdef}
\end{equation}
where $p_{ij}(\hat{k}) =[ \delta_{ij} - \hat{k}_{i} \hat{k}_{j}]$ is the traceless projector. The field operators 
(\ref{oneE}) and (\ref{twoE}) consist of a positive and of a negative frequency part, i.e. 
$\hat{h}_{ij}(x) = \hat{h}^{(+)}_{ij}(x) + \hat{h}^{(-)}_{ij}(x)$ with $ \hat{h}^{(+)\,\,\dagger}_{ij}(x) =  \hat{h}^{(-)}_{ij}(x)$.
If $| \mathrm{vac} \rangle$ is the state that minimizes the tensor Hamiltonian when all the modes are inside 
the effective horizon (for instance at the onset of inflation) the operator $\hat{h}^{(+)}_{ij}(x)$ annihilates 
the vacuum [i.e. $ \hat{h}^{(+)}_{ij}(x) \,| \mathrm{vac} \rangle=0$ and $\langle \mathrm{vac} | \,\hat{h}^{(-)}_{ij}(x) =0$].
The two-point functions associated with $\hat{h}_{ij}$ and  $\hat{h}^{\,\,\prime}_{ij}$ are therefore given by:
\begin{eqnarray}
\langle \mathrm{vac} | \hat{h}_{ij}(\vec{x}, \tau) \hat{h}_{ij}(\vec{x} + \vec{r}, \tau) | \mathrm{vac}\rangle 
&=& \int \,d \ln{k} \,\,P_{T}(k,\tau) \,\,j_{0}(k r), 
\label{fourEa}\\
\langle \mathrm{vac} | \hat{h}^{\,\,\prime}_{ij}(\vec{x}, \tau) \hat{h}^{\,\,\prime}_{ij}(\vec{x} + \vec{r}, \tau) | \mathrm{vac}\rangle 
&=& \int \,d \ln{k} \,\,Q_{T}(k,\tau) \,\,j_{0}(k r),
\label{fourEb}
\end{eqnarray}
and $j_{0}(k r)$ are spherical Bessel functions of zeroth order \cite{twentyfour,twentyfive}; $P_{T}(k,\tau)$ is the standard tensor power spectrum while $Q_{T}(k,\tau)$ is usually not discussed 
but its presence is essential in the present context:
\begin{equation}
P_{T}(k,\tau) = \frac{4 \ell_{P}^2}{\pi^2} \,\,k^3 \,\,\bigl| F_{k}(\tau) \bigr|^2, \qquad Q_{T}(k,\tau) = \frac{4 \ell_{P}^2}{\pi^2} \,\,k^3 \,\,\bigl| G_{k}(\tau) \bigr|^2.
\label{fiveE}
\end{equation}
In Eqs. (\ref{fourEa})--(\ref{fourEb}) the expectation values have been computed over the vacuum state. The averages of the field operators can also be obtained directly in Fourier space 
from the corresponding Fourier transforms; by representing  the quantum operators 
in Fourier space as 
\begin{equation}
\hat{h}_{ij}(\vec{k},\tau) = \frac{1}{(2\pi)^{3/2}} \int d^{3} x\, e^{i \vec{k}\cdot\vec{x}} \, \hat{h}_{i\, j}(\vec{x}, \tau),\qquad 
\hat{h}^{\,\,\prime}_{ij}(\vec{k},\tau) = \frac{1}{(2\pi)^{3/2}} \int d^{3} x\, e^{i \vec{k}\cdot\vec{x}} \, \hat{h}_{i\, j}^{\,\,\prime}(\vec{x}, \tau),
\label{eightE}
\end{equation}
the explicit expressions of $\hat{h}_{ij}(\vec{k},\tau)$ and of $\hat{h}^{\,\,\prime}_{ij}(\vec{k},\tau) $ follow
from Eqs. (\ref{oneE}) and (\ref{twoE}) so that the corresponding expectation values are 
\begin{eqnarray}
\langle \hat{h}_{i\, j}(\vec{k},\,\tau) \hat{h}_{m\, n}(\vec{p},\,\tau) \rangle &=& \frac{2 \pi^2}{k^3} \,\, P_{T}(k,\tau) \, 
\delta^{(3)}(\vec{k} + \vec{p}) \,\,{\mathcal S}_{i\,j\,m\,n},
\label{sixE}\\
\langle \hat{h}_{i\, j}^{\, \prime}(\vec{k},\,\tau) \hat{h}_{m\, n}^{\, \prime}(\vec{p},\,\tau) \rangle &=& \frac{2 \pi^2}{k^3} \,\, Q_{T}(k,\tau) \, \delta^{(3)}(\vec{k} + \vec{p}) \,\,{\mathcal S}_{i\,j\,m\,n}.
\label{sevenE}
\end{eqnarray}
The expressions of Eqs. (\ref{sixE}) and (\ref{sevenE}) hold for quantum mechanical operators but 
can be easily viewed as classical expectation values of isotropic random fields, as we shall discuss in section \ref{sec5}.
The expectation values of the energy density and of the pressures 
in the different parametrizations examined above will be computed in the remaining part of this section by using the following notations\footnote{In the Landau-Lifshitz parametrization we have to add also the spectral density corresponding to the shifted pressure ${\mathcal P}_{gw}^{(L)}(k,\tau)$ (see Eq. (\ref{sixG})) that 
is, in some sense, the true pressure term arising from the second order fluctuation of the Bianchi identity. 
The contribution of the shifted pressure has been sometimes interpreted as an effective bulk viscosity of the relic gravitons \cite{ten} but this suggestion shall not be pursued here.}:
\begin{equation}
\overline{\rho}_{gw}^{(X)} =  \langle \mathrm{vac} | \hat{\rho}_{gw}^{(X)} | \mathrm{vac} \rangle\, \qquad \overline{p}_{gw}^{(X)} =  \langle \mathrm{vac} | \hat{p}_{gw}^{(X)} | \mathrm{vac} \rangle,
\label{fourXX}
\end{equation}
where $X =F,\, L,\,B$. From Eq. (\ref{fourXX}) it is also practical to introduce the 
spectral energy density and the spectral pressure defined as 
\begin{equation}
\rho^{(X)}_{gw}(k,\tau) = \frac{d \overline{\rho}^{(X)}_{gw}}{d \ln{k}}, \qquad p^{(X)}_{gw}(k,\tau) = \frac{d \overline{\rho}^{(X)}_{gw}}{d \ln{k}}.
\label{oneL}
\end{equation}
The quantum averaging implies a correct ordering of the operators: for instance the quantum version of the classical expression 
$2 {\mathcal H} \partial_{\tau} h_{k\ell}\, h^{k\ell}$ reads, as usual, 
${\mathcal H}( \partial_{\tau} \hat{h}_{k\ell}\, \hat{h}^{k\ell} +  \hat{h}_{k\ell}\, \partial_{\tau}\hat{h}^{k\ell})$.
The choice of $|\mathrm{vac} \rangle$ is not mandatory:  if the vacuum state is replaced by some other initial state the present considerations apply in the same way provided the {\em same} initial state is used for all the expectation values in the various descriptions. 

\subsection{The effective energy momentum pseudo-tensor}
In spite of the specific parametrization of the energy-momentum pseudo-tensor, it is a general 
property of the quantum mechanical expectation values that the averages of the anisotropic stresses 
and of the total derivatives are all vanishing: 
\begin{eqnarray}
&& \langle \, \mathrm{vac} | \hat{\Pi}_{i}^{(F)\, j} | \mathrm{vac} \rangle = \langle \, \mathrm{vac} | \hat{\Pi}_{i}^{(B)\, j} | \mathrm{vac} \rangle = \langle \, \mathrm{vac} | \hat{\Pi}_{i}^{(B)\, j} | \mathrm{vac} \rangle =0,
\label{oneF}\\
&& \langle \partial_{i} \biggl[ \hat{h}_{k\,\ell} \partial^{i} \hat{h}^{k\,\ell}\biggr] \rangle = \langle \partial_{i} \biggl[ \hat{h}_{k\,\ell} \partial^{\ell} \hat{h}^{k\,i}\biggr] \rangle =0. 
\label{twoF}
\end{eqnarray}
Similarly the expectation values of the three-divergences of the energy fluxes vanish, i.e. 
\begin{equation}
\langle \vec{\nabla} \cdot \vec{S}^{(F)} \rangle =\, \langle \vec{\nabla} \cdot \vec{S}^{(L)} \rangle =\,\langle \vec{\nabla} \cdot \vec{Q}^{(L)} \rangle=\,
 \langle \vec{\nabla} \cdot \vec{S}^{(B)} \rangle =\,\,0.
\label{threeF}
\end{equation}
While the results of Eqs. (\ref{oneF}) and (\ref{twoF})--(\ref{threeF})  hold for all the cases examined above, 
the spectral energy densities and the spectral pressures of Eq. (\ref{oneL}) are all different.
In the case $X=F$ (see Eq. (\ref{oneL})) the spectral energy density and the spectral pressure become:
\begin{eqnarray} 
\rho_{gw}^{(F)}(k,\tau) &=& \frac{d \overline{\rho}_{gw}^{(F)}}{d \ln{k}} = \frac{1}{8 \ell_{P}^2 a^2} \biggl[ Q(k,\tau) + k^2 P_{T}(k,\tau) \biggr],
\label{sixF}\\
p_{gw}^{(F)}(k,\tau) &=& \frac{d \overline{p}_{gw}^{(F)}}{d \ln{k}} = \frac{1}{8 \ell_{P}^2 a^2} \biggl[ Q(k,\tau) - \frac{k^2}{3} P_{T}(k,\tau) \biggr].
 \label{sevenF}
\end{eqnarray}
Recalling the explicit form of the power spectra given in Eq. (\ref{fiveE}), $\rho_{gw}^{(F)}(k,\tau)$ and $p_{gw}^{(F)}(k,\tau)$ 
can also be expressed as:
\begin{eqnarray}
\rho_{gw}^{(F)}(k,\tau) &=& \frac{k^3}{2 \pi^2 a^2} \biggl[ k^2 \bigl| F_{k}(\tau)\bigr|^2 +   \bigl| G_{k}(\tau)\bigr|^2 \biggr],
\label{eightF}\\
p_{gw}^{(F)}(k,\tau) &=& \frac{k^3}{2 \pi^2 a^2} \biggl[ - \frac{k^2}{3} \bigl| F_{k}(\tau)\bigr|^2 +   \bigl| G_{k}(\tau)\bigr|^2 \biggr].
\label{nineF}
\end{eqnarray}
When the typical frequencies of the gravitons 
are much larger than the rate of variation of the geometry,  Eq. (\ref{threeE}) implies that 
$|G_{k}(\tau)|^2 = [ k^2 + {\mathcal H}^2  + {\mathcal O}({\mathcal H}^2/k^2)] |F_{k}(\tau)|^2$;
therefore in the limit $k \gg {\mathcal H}$  the effective barotropic index computed from 
Eqs. (\ref{eightF})--(\ref{nineF}) becomes:
 \begin{equation}
 \lim_{k\gg {\mathcal H}} \frac{p_{gw}(k,\tau)}{\rho_{gw}(k,\tau)} = \frac{1}{3} \biggl( 1 + \frac{{\mathcal H}^2}{k^2} \biggr).
 \label{tenF}
 \end{equation}
According to Eq. (\ref{tenF}), when the modes are inside the Hubble radius the barotropic index 
of the relic gravitons coincides approximately with $1/3$ to leading order in ${\mathcal H}^2/k^2 \ll 1$ (i.e. $k \tau \gg 1$).  
In the opposite limit (i.e. $k^2/{\mathcal H}^2\ll 1$) the frequency of the waves is much smaller than the rate of variation 
of the geometry and Eq. (\ref{threeE}) can be solved by iteration:
 \begin{eqnarray}
 F_{k}(\tau) &=& F_{k}(\tau_{ex}) + G_{k}(\tau_{ex}) \int_{\tau_{ex}}^{\tau} \frac{a_{ex}^2}{a^2(\tau_{1})} \, d\tau_{1} 
 - k^2 \int_{\tau_{ex}}^{\tau} \frac{ d\tau_{2}}{a^2(\tau_{2})} \int_{\tau_{ex}}^{\tau_{2}} F_{k}(\tau_{1}) \, d\tau_{1}
 \label{elevenF}\\
 G_{k}(\tau) &=& \biggl(\frac{a_{ex}}{a}\biggr)^2 G_{k}(\tau_{ex}) - \frac{k^2}{a^2} \int_{\tau_{ex}}^{\tau} F_{k}(\tau_{1}) d\tau_{1},
 \label{twelveF}
 \end{eqnarray}
where $\tau_{ex}$ is the conformal time corresponding to the exit of a given wavelength from the Hubble radius (i.e.
$k \tau_{ex} = {\mathcal O}(1)$).
From Eqs. (\ref{elevenF}) and (\ref{twelveF}) we have that in the limit $k \ll {\mathcal H}$ 
 \begin{eqnarray}
\rho_{gw}^{(F)}(k,\tau) &=& \frac{k^3}{2 \pi^2 a^2} \biggl[ k^2 \bigl| F_{k}(\tau_{ex})\bigr|^2 +  \biggl(\frac{a_{ex}}{a}\biggr)^4 \bigl| G_{k}(\tau_{ex})\bigr|^2 \biggr],
\label{thirteenF}\\
p_{gw}^{(F)}(k,\tau) &=& \frac{k^3}{2 \pi^2 a^2} \biggl[ - \frac{k^2}{3} \bigl| F_{k}(\tau_{ex})\bigr|^2 + \biggl(\frac{a_{ex}}{a}\biggr)^4 \bigl| G_{k}(\tau_{ex})\bigr|^2 \biggr].
\label{fourteenF}
\end{eqnarray}
If the background expands, the terms proportional to $\bigl| G_{k}(\tau_{ex})\bigr|^2$ quickly becomes 
negligible and the effective barotropic index goes to $-1/3$; conversely if the background contracts 
the first term becomes  subleading and the effective barotropic index tends asymptotically towards 
$1$:
\begin{eqnarray}
w_{gw} &=& \frac{p_{gw}^{(F)}(k,\tau) }{\rho_{gw}(k,\tau)} \to - \frac{1}{3} , \qquad  k^2 \bigl| F_{k}(\tau_{ex})\bigr|^2 \gg   \biggl(\frac{a_{ex}}{a}\biggr)^4 \bigl| G_{k}(\tau_{ex})\bigr|^2,
\label{fifteenF}\\
w_{gw} &=& \frac{p_{gw}^{(F)}(k,\tau) }{\rho_{gw}(k,\tau)} \to 1 , \qquad  k^2 \bigl| F_{k}(\tau_{ex})\bigr|^2 \ll   \biggl(\frac{a_{ex}}{a}\biggr)^4 \bigl| G_{k}(\tau_{ex})\bigr|^2.
\label{sixteenF}
\end{eqnarray}
All in all we can then say that when the typical frequency of the gravitons exceeds the rate of variation of the geometry 
(i.e. $k \gg {\mathcal H}$)  the high-frequency gravitons behave as a perfect relativistic fluid and the barotropic index is 
$1/3$. In the opposite 
limit  (i.e. $k \ll {\mathcal H}$) the effective barotropic index 
becomes $-1/3$ if the background expands while it becomes $1$ if the background contracts. 
Equation (\ref{sevenB}) can also be averaged term by term and the result will be an evolution 
equation for the expectation values of the energy density and of the pressure, i.e. 
\begin{equation}
\partial_{\tau} \overline{\rho}^{(F)}_{gw} + 3 {\mathcal H} [ \overline{\rho}^{(F)}_{gw} + \overline{p}^{(F)}_{gw} ] =0,
\label{seventeenF}
\end{equation}
where the contribution of the energy flux originally present in Eq. (\ref{sevenB}) 
disappears because of Eq. (\ref{threeF}).

\subsection{The Landau-Lifshitz pseudo-tensor}
The quantum averaging of the Landau-Lifshitz pseudo-tensor leads to the same 
results of the effective energy-momentum tensor in the high-frequency limit but the results 
are sharply different when the frequency is smaller than the Hubble rate. Using the same 
notations of Eq. (\ref{fourXX})  we have that, in this case, the spectral distributions 
are:
\begin{eqnarray}
\rho_{gw}^{(L)}(k,\tau)  &=&  \frac{k^3}{2 \pi^2 a^2} \biggl[ k^2 \bigl| F_{k}(\tau)\bigr|^2 +   \bigl|G_{k}(\tau)\bigr|^2 + 4 {\mathcal H} ( G_{k} \, F_{k}^{*} + G_{k}^{*} F_{k})\biggr],
\label{fourG}\\
p_{gw}^{(L)}(k,\tau) &=& \frac{k^3}{6 \pi^2 a^2} \biggl[ 7 k^2  \bigl| F_{k}(\tau)\bigr|^2 - 5 \bigl| G_{k}(\tau)\bigr|^2 \biggr],
\label{fiveG}\\
{\mathcal P}_{gw}^{(L)}(k,\tau) &=& \frac{k^3}{6 \pi^2 a^2} \biggl[ 7 k^2  \bigl| F_{k}(\tau)\bigr|^2 - 5 \bigl| G_{k}(\tau)\bigr|^2 + 4 \biggl( {\mathcal H} - \frac{{\mathcal H}^{\prime}}{{\mathcal H}}\biggr)(G_{k} F_{k}^{*} + G_{k}^{*} F_{k})\biggr].
\label{sixG}
\end{eqnarray}
Equations (\ref{fiveG}) and (\ref{sixG}) give the pressure and the shifted pressure respectively;
note that ${\mathcal P}_{gw}^{(L)}$ enters the conservation 
equation obeyed by the mean values:
\begin{equation}
\partial_{\tau} \overline{\rho}^{(L)}_{gw} + 3 {\mathcal H} [ \overline{\rho}^{(L)}_{gw} + \overline{{\mathcal P}}^{(L)}_{gw} ] =0,
\label{sevenG}
\end{equation}
and it coincides with $p_{gw}^{(L)}$ when ${\mathcal H}^2 = {\mathcal H}^{\prime}$, i.e. in the case of an exact de Sitter expansion.
Inside the Hubble radius (i.e. for $k \gg {\mathcal H}$), Eqs. (\ref{fourG}), (\ref{fiveG}) and (\ref{sixG}) imply:
\begin{equation}
\lim_{k \gg {\mathcal H}} \frac{p_{gw}^{(L)}(k,\tau) }{\rho_{gw}^{(L)}(k,\tau)} = \frac{{\mathcal P}_{gw}^{(L)}(k,\tau) }{\rho_{gw}^{(L)}(k,\tau)} =  \frac{1}{3} \biggl( 1 + \frac{{\mathcal H}^2}{k^2} \biggr),
\label{eightG}
\end{equation}
 We can then conclude, as expected, that the result (\ref{eightG}) coincides with Eq. (\ref{tenF})
(following, in turn, from Eqs. (\ref{sixF})--(\ref{sevenF})). In the opposite physical regime (i.e. when $k \ll {\mathcal H}$), however,  
 quantitative conclusions cannot be deduced in general terms and the best strategy will then be (see section \ref{sec4}) to analyze a number of specific examples 
 by explicitly computing the spectral energies and pressures. This discussion will allow for a fair comparison among the results of
 Eqs. (\ref{fourG})--(\ref{sixG}) and Eqs. (\ref{sixF})--(\ref{sevenF}) in the low-frequency domain where 
$k \ll {\mathcal H}$. 

\subsection{The Brill-Hartle scheme and the quantum averaging}
The quantum average of the Brill-Hartle-Isaacson results of Eqs. (\ref{rhoB}) and (\ref{pB})
leads to the following spectral energy and pressure: 
\begin{equation}
\rho_{gw}^{(B)}(k,\tau)  =  \frac{k^3}{\pi^2 a^2}  \bigl|G_{k}(\tau)\bigr|^2 ,\qquad 
p_{gw}^{(B)}(k,\tau)  =  \frac{k^5}{3\pi^2 a^2}  \bigl|F_{k}(\tau)\bigr|^2.
\label{fourH}
\end{equation}
As usual in the limit $k \gg {\mathcal H}$ we have that $p_{gw}^{(B)}(k,\tau)/\rho_{gw}^{(B)}(k,\tau) \to 1/3$ since
 $\bigl|G_{k}(\tau)\bigr|^2 = (k^2 + {\mathcal H}^2) \bigl|F_{k}(\tau)\bigr|^2$ when the corresponding wavelengths 
 are shorter than the Hubble radius. In the opposite limit, however, Eq. (\ref{fourH}) leads to a bizarre result: 
 the energy density is asymptotically vanishing (i.e. $\rho_{gw}^{(B)}(k,\tau) \to 0$) and the spectral pressure 
 becomes much larger than  $\rho_{gw}^{(B)}(k,\tau)$ (i.e. $p_{B}(k,\tau)/\rho_{B}(k,\tau) \gg 1$) and it is formally divergent. 
 For contracting backgrounds the opposite is true: the spectral pressure gets progressively more negligible 
(i.e. $p_{gw}^{(B)}(k,\tau) \to 0$) so that $p_{B}(k,\tau)/\rho_{B}(k,\tau) \ll 1$.
These apparent inconsistencies (further explored in the concrete examples of sections \ref{sec4} and \ref{sec5})
 can be expected as long as  Brill-Hartle average automatically selects all the modes that 
are inside the Hubble radius and it is therefore not surprising 
that they lead to quantitative ambiguities when the wavelengths exceed the Hubble radius. It is  possible to obtain a covariant expression that also applies beyond the Hubble radius. In this case, however, 
the energy density and pressure do not follow from the Brill-Hartle scheme. To 
prove this statement let us neglect, for simplicity, the potential sources  and write the full 
second-order action in the covariant case:
\begin{equation}
S_{cov} = \frac{1}{8 \ell_{P}^2}\int \, \, \sqrt{- \overline{g}}\,\, d^{4} x\,\, \biggl[ \overline{g}^{\alpha\beta}\,\,\overline{\nabla}_{\alpha} \widetilde{h}_{\mu\nu} \overline{\nabla}_{\beta} \widetilde{h}^{\mu\nu} + 2 \overline{R}^{\,\,\gamma\,\,\,\,\,\,\,\,\alpha}_{\,\,\,\,\,\mu\nu} \, \widetilde{h}_{\gamma\alpha} \widetilde{h}^{\mu\nu} \biggr],
\label{fiveH}
\end{equation}
where the conditions of Eq. (\ref{oneD}) have been consistently imposed. 
By extremizing the action (\ref{fiveH}) with respect to the variation of  $ \widetilde{h}^{\mu\nu}$ we obtain the evolution equation  for the covariant tensor amplitude
\begin{equation}
\overline{\nabla}_{\alpha} \overline{\nabla}^{\alpha} \widetilde{h}_{\mu\nu} 
- 2\,\,\overline{R}^{\,\,\gamma}_{\,\,\,\,\,\mu\nu\alpha} \, \widetilde{h}_{\gamma}^{\,\,\,\,\alpha} =0.
\label{sixH}
\end{equation}
If we now consider the background metric and the fluctuating amplitude 
as independent variables the energy-momentum tensor following from Eq. (\ref{fiveH})
reads:
\begin{eqnarray}
T_{\mu\nu}^{(gw)} &=& \frac{1}{4\ell_{P}^2} \biggl\{ \overline{\nabla}_{\mu} \widetilde{h}_{\alpha\beta}  \overline{\nabla}_{\nu} \widetilde{h}^{\,\,\alpha\beta}  +  \overline{\nabla}_{\alpha} \widetilde{h}_{\mu\beta} \,\, \overline{\nabla}^{\alpha} \widetilde{h}^{\,\,\,\,\beta}_{\nu} +  \overline{\nabla}_{\alpha} \widetilde{h}_{\nu\beta}\,\,  \overline{\nabla}^{\alpha} \widetilde{h}^{\,\,\,\,\beta}_{\mu} 
\nonumber\\
&+& 2\,\, \overline{R}^{\gamma}_{\,\,\,\,\,\,\mu \rho \alpha} \, \widetilde{h}_{\gamma}^{\,\,\alpha} \, \widetilde{h}_{\nu}^{\,\,\rho}
+ 2 \,\,\overline{R}^{\,\,\gamma}_{\,\,\,\,\,\, \nu \rho \alpha} \, \widetilde{h}_{\gamma}^{\,\,\alpha} \, \widetilde{h}^{\,\,\,\rho}_{\mu}
\nonumber\\
&-& \frac{1}{2} \overline{g}_{\mu\nu} \biggl[ \overline{\nabla}_{\rho} \widetilde{h}_{\alpha\beta} \overline{\nabla}^{\rho} \widetilde{h}_{\alpha\beta} + 2 \overline{R}^{\,\,\gamma\,\,\,\,\,\,\,\,\rho}_{\,\,\,\,\,\alpha\beta} \, \widetilde{h}_{\gamma\rho} \,\,\,\widetilde{h}^{\alpha\beta} \biggr]\biggr\}.
\label{sevenH}
\end{eqnarray}
If we now apply the tenets of the Brill-Hartle procedure \cite{four} the covariant gradients average out to zero. 
Therefore we can flip the covariant derivative from one amplitude to the other. If we do this with the terms inside the squared bracket  
of Eq. (\ref{sevenH}) we can obtain terms like $\widetilde{h}^{\alpha\beta} \overline{\nabla}_{\rho} \overline{\nabla}^{\rho} \widetilde{h}_{\alpha\beta}$. Using Eq. (\ref{sixH}) all these terms will produce various Riemann tensors 
that will be neglected so that, at the very end, the only term surviving the average will be the first contribution 
of Eq. (\ref{sevenH}), i.e. 
\begin{equation}
T_{\mu\nu}^{(gw)} = {\mathcal B}_{\mu\nu}  = \frac{1}{4 \ell_{P}^2} \langle \overline{\nabla}_{\mu} \widetilde{h}_{\alpha\beta}  \overline{\nabla}_{\nu} \widetilde{h}^{\alpha\beta} \rangle_{BH}  = \frac{1}{4 \ell_{P}^2} \langle \partial_{\mu} h_{ij}  \partial_{\nu} \, h^{i j} \rangle_{BH},
\label{eightH}
\end{equation}
where the Brill-Hartle average has been intentionally indicated to clarify the origin of the term. 
The second equality in Eq. (\ref{eightH}) 
follows by making explicit the covariant derivatives and by appreciating that, within the present 
definitions, $\widetilde{h}_{i\, j} = - a^2 h_{ij}$ while the other components of $\widetilde{h}_{\mu\nu}$ vanish.
Equation (\ref{eightH}) coincides with Eq. (\ref{threeD}) and it shows that the Brill-Hartle average 
effectively neglects all the terms that are relevant beyond the Hubble radius. 
A result applicable beyond the Hubble radius follows from 
 Eq. (\ref{fiveH}) but without imposing the Brill-Hartle averaging: if we 
 use Eq. (\ref{fiveH}) and express it in the conformally flat case 
 (i.e. $\overline{g}_{\mu\nu} = a^2 \eta_{\mu\nu}$ and $ \widetilde{h}_{ij} = - a^2 h_{ij}$)
 we obtain, after a lengthy but straightforward calculation, the same expression 
 of the effective energy-momentum tensor ${\mathcal F}_{\mu\nu}$ 
 given in Eqs. (\ref{rhoF})--(\ref{pF}).
 
\renewcommand{\theequation}{4.\arabic{equation}}
\section{The spectral energy density}
\setcounter{equation}{0}
\label{sec4}
The salient properties of the different pseudo-tensors in the case of expanding backgrounds are summarized, 
for the sake of conciseness,  in Tab. \ref{TABLE1}. While the basic features of ${\mathcal F}_{\mu\nu}$ have been deduced
without assuming any specific evolution of the background, the physical properties of ${\mathcal L}_{\mu\nu}$ 
in the long wavelength limit demand a more concrete analysis of some specific examples. Table \ref{TABLE1}  also suggests 
that the Brill-Hartle results are only applicable in the high-frequency regime and they must be otherwise completed 
by the full expression of the covariant energy-momentum tensor (see Eq. (\ref{sevenH}) which coincides 
with ${\mathcal F}_{\mu\nu}$ in the conformally flat case. 
\begin{table}[!ht]
\begin{center}
\begin{tabular}{||c|c|c|c|c||}
\hline
\hline
\rule{0pt}{4ex} pseudo-tensor& $w_{gw}$ ($k\tau\gg 1$)  & $w_{gw}$ ($k\tau \ll 1$) & $\rho_{gw}^{(X)}$ ($k\tau \gg 1$) & $\rho_{gw}^{(X)}$($k\tau \ll 1$) \\
\hline
\hline
${\mathcal F}_{\mu\nu}$& $1/3$ & $-\,1/3$  & $\rho_{gw}^{(F)} \geq 0$ & $\rho_{gw}^{(F)} \geq 0$ \\
${\mathcal L}_{\mu\nu}$& $1/3$ & undetermined & $\rho_{gw}^{(L)}\geq 0 $ & undetermined \\
${\mathcal B}_{\mu\nu}$ & $1/3$ & not applicable & $\rho_{gw}^{(B)}  \geq 0$ & not applicable \\
\hline
\hline
\end{tabular}
\caption{Summary of the salient properties of the different pseudo-tensors in the case of expanding backgrounds; 
$w_{gw}$ denotes the ratio of the spectral pressure and of the spectral energy density in the different cases.}
\label{TABLE1}
\end{center}
\end{table}
A table similar to Tab. \ref{TABLE1} can be compiled in the case of contracting 
backgrounds\footnote{For instance $-1/3$ must be substituted 
by $1$, as the general considerations of Eq. (\ref{sixteenF}) demonstrate (see also Ref. \cite{ten} 
for some explicit example).} but in what follows the ambiguities of Tab. \ref{TABLE1} will be addressed 
mainly in the case of expanding backgrounds. When the wavelengths are all inside the Hubble radius, the frequency range 
of the spectrum roughly ranges between the aHz and $100$ MHz (i.e. $10^{-18}$ Hz and $10^{8}$ Hz).
In backreaction problems (see e.g. \cite{nine,ten}) the averaged energy density and pressure
beyond the Hubble radius are determined by integrating the spectral energy density and the spectral pressure 
over $d \ln{k}$ between the fixed extrema $k_{ex}$ and $k_{re}$ 
 corresponding to the wavelengths that exit and reenter the Hubble radius.

\subsection{The expanding branch of the de Sitter space-time}
If de Sitter space is exact (i.e. in the absence of slow-roll corrections) the scale 
factor is given by $a_{i}(\tau) = (- \tau_{1}/\tau)$ with ${\mathcal H}= - 1/\tau$; the 
scalar modes are absent but the propagating tensors are characterized by 
the following mode function:
\begin{equation}
F_{k}(\tau) = \frac{1}{\sqrt{2 k} \, a(\tau)} \biggl( 1 - \frac{i }{k \tau} \biggr) \, e^{- i \, k \tau}, \qquad \tau \leq - \tau_{1},
\label{MF}
\end{equation}
where the boundary conditions follow from Eq. (\ref{threeE}). The spectral energy 
density is in general a function of $k$ and $\tau$ but 
if we introduce $x = | k \tau |$ the spectral energy and pressure are both 
functions of the single dimensionless variable $x$:
\begin{equation}
\rho_{gw}^{(F)}(x) = \frac{H_{1}^4}{4 \pi^2} \biggl[ x^2 (2 x^2 +1)\biggr], \qquad 
p_{gw}^{(F)}(x) = \frac{H_{1}^4}{12 \pi^2} \biggl[ x^2 (2 x^2 -1)\biggr], 
\label{twoL}
\end{equation}
where $H_{1} a_{1} \equiv H_{1} = 1/\tau_{1}$ [recall that $a_{1} = a(- \tau_{1}) =1$]. According to Eq. (\ref{twoL})  
the spectral energy is always positive semi-definite and that the effective barotropic index interpolates 
between $-1/3$ (when $x \ll 1$) and $1/3$ (when $x \gg 1$). Both results have been anticipated 
in Tab. \ref{TABLE1} on the basis of the general considerations of section \ref{sec3}. Using then Eq. (\ref{twoL})
we have:
\begin{equation}
\rho_{gw}^{(F)}(x) \geq 0, \qquad \lim_{x \gg 1} \frac{p_{gw}^{(F)}(x)}{\rho_{gw}^{(F)}(x)} = \frac{1}{3}, 
\qquad  \lim_{x \ll 1} \frac{p_{gw}^{(F)}(x)}{\rho_{gw}^{(F)}(x)} = - \frac{1}{3},
\label{threeL}
\end{equation}
so that Eq. (\ref{threeL}) agrees exactly with Eqs. (\ref{tenF}) and (\ref{sixteenF}) since 
the limit $ x \ll 1$ corresponds to those frequencies that are smaller than the 
rate of variation of the geometry while in the regime $ x \gg 1$ the frequencies exceed  
${\mathcal H}$.

In the case of  ${\mathcal L}_{\mu\nu}$ the same analysis leading to Eq. (\ref{threeL}) solves some of the ambiguities 
listed in Tab. \ref{TABLE1}. In particular, 
using Eq. (\ref{MF}) into Eqs.  (\ref{fourG})--(\ref{fiveG}) (recall also Eqs. (\ref{rhoL})--(\ref{pL})) the
spectral energy density and the spectral pressure are:
\begin{equation}
\rho_{gw}^{(L)}(x) = \frac{H_{1}^4}{4 \pi^2} \biggl[x^2 (2 x^2 - 7)\biggr], \qquad 
p_{gw}^{(L)}(x) = {\mathcal P}_{gw}^{(L)}(x)  = \frac{H_{1}^4}{12 \pi^2} \biggl[ x^2 (2 x^2 + 7)\biggr]. 
\label{fourL}
\end{equation}
According to Eq. (\ref{fourL}) spectral energy density does not have a definite sign since it is positive 
inside the Hubble radius but negative outside:
\begin{equation}
\lim_{x \gg 1} \rho_{gw}^{(L)}(x) =  \frac{H_{1}^4}{2 \pi^2} x^4, \qquad \lim_{x \ll 1} \rho_{gw}^{(L)}(x) =  - \frac{7 H_{1}^4}{4 \pi^2} x^2,
\label{fiveL}
\end{equation}
in agreement with previous results \cite{eight,nine,ten}.  Since in the de Sitter case ${\mathcal H}^2 = {\mathcal H}^{\prime}$ we also have that the pressure and the shifted pressure coincide, i.e. $p_{gw}^{(L)}(x) = {\mathcal P}_{gw}^{(L)}(x) $; the effectice barotropic index is then given by 
\begin{equation}
\lim_{x \gg 1} \frac{{\mathcal P}_{gw}^{(L)}(x)}{\rho_{gw}^{(L)}(x)} = \frac{1}{3}, 
\qquad  \lim_{x \ll 1} \frac{{\mathcal P}_{gw}^{(L)}(x)}{\rho_{gw}^{(L)}(x)} = - \frac{1}{3},
\label{sixL}
\end{equation}
which is formally the same result of Eq. (\ref{threeL}) with the difference that the signs are inverted:
the averaged energy density is negative while the corresponding pressure is positive. Taken at face 
value the result of Eq. (\ref{sixL}) violates the weak energy conditions but it is difficult to attribute a 
profound physical meaning to this occurrence as long as there exist other pseudo-tensors (like ${\mathcal F}_{\mu\nu}$)
not violating the weak energy condition. 

In the Brill-Hartle case Eqs. (\ref{rhoB})--(\ref{pB}) and (\ref{fourH})  imply that the corresponding spectral energy density 
and pressure are 
\begin{equation}
\rho_{gw}^{(B)}(x) = \frac{H_{1}^4}{2 \pi^2} x^4, \qquad 
p_{gw}^{(B)}(x) = \frac{ H_{1}^4}{6 \pi^2} \biggl[ x^2 ( x^2 + 1)\biggr]. 
\label{sevenL}
\end{equation}
According to Eq. (\ref{sevenL}) the energy density is positive semidefinite but the effective barotropic 
index diverges in the limit $x \to 0 $:
\begin{equation}
\rho_{gw}^{(B)}(x) \geq 0, \qquad \lim_{x \gg 1} \frac{p_{gw}^{(B)}(x)}{\rho_{gw}^{(B)}(x)} = \frac{1}{3}, 
\qquad  \lim_{x \ll 1} \frac{p_{gw}^{(B)}(x)}{\rho_{gw}^{(B)}(x)} \simeq \frac{1}{3 x^4}. 
\label{eightL}
\end{equation}
Equation (\ref{eightL}) confirms the conclusion of section \ref{sec3} where it has been 
shown, on a general ground, that the Brill-Hartle approach selects a priori only the wavelengths inside 
the Hubble radius and it gives the same result of all the other strategies only in this physical domain.

\subsection{The expanding de Sitter background matched to radiation}
If  the mode function normalized during the de Sitter phase 
but it evolves through radiation the modes not only exit the Hubble radius but they can also reenter. The 
spectral energy density and pressure can then be expressed in terms of two dimensionless variables\footnote{Note that, unlike the pure de Sitter case, we find it more convenient to define $x = k\tau$ (and not $x = |k\tau|$ as in the previous case).} i.e. $x = k\tau$ and $y = k \tau_{1}$. The scale factor for $\tau \geq -\tau_{1}$ is linear (as in the case of a radiation-dominated 
regime) $a_{r}(\tau) = (\tau + 2 \tau_{1})/\tau_{1}$ and it is 
continuously matched to $a_{i}(\tau) = (- \tau_{1}/\tau)$. The scale 
factor and the its rate of variation are both continuous in $-\tau_{1}$, i.e. 
$a_{i}(-\tau_{1}) = a_{r}(-\tau_{1})$ and ${\mathcal H}_{i}(-\tau_{1})= 
{\mathcal H}_{r}(-\tau_{1})$. With these precisions we have that for $\tau \geq - \tau_{1}$ the mode functions are 
given by 
\begin{equation}
F_{k}(\tau) = \frac{1}{ \sqrt{2 k} \,a_{r}(\tau)} \biggl[ c_{+}(k, \tau_{1}) e^{ - i k (\tau+ 2 \tau_{1}) }+ c_{-}(k, \tau_{1}) e^{  i k (\tau+ 2 \tau_{1})}\biggr], 
\label{oneM}
\end{equation}
where ${\mathcal H} = 1/(\tau + 2 \tau_{1})$ and  $G_{k}= F_{k}^{\prime}$.  The complex coefficients $c_{\pm}(k,\tau_{1})$ appearing in Eq. (\ref{oneM}) 
obey $|c_{+}(k,\tau_{1})|^2 - |c_{-}(k,\tau_{1})|^2 =1$ and are given by:
\begin{equation}
c_{+}( y) = \frac{e^{ 2 \,i\, y} ( 2 y^2 + 2 \, i\, y-1)}{2 y^2}, \qquad c_{-}( y) = \frac{1}{2 y^2},
\label{twoM}
\end{equation}
where we introduced the dimensionless variable $y = k \tau_{1}$.  The spectral energy density 
and the pressure have an exact expression which is however not so revealing. For instance 
in the case of ${\mathcal F}_{\mu\nu}$ we have:
\begin{eqnarray}
\rho_{gw}^{(F)}(x,y) &=& \frac{H_{1}^4\, y^3}{8 \pi ^2 (x+2 y)^6} \biggl[\left(2 y^4+1\right) \left(2 x^2+8 x y+8 y^2+1\right)
\nonumber\\
&+&\left(4 x y^2-2 x+8 y^3-2 y\right)
   \sin{2 (x+y)} - \left(4 x y+6 y^2+1\right) \cos{2 (x+y)}\biggr],
\nonumber\\
 p_{gw}^{(F)}(x,y) &=& \frac{ H_{1}^4 y^3}{24 \pi ^2 (x+2 y)^6} \biggl[\left(2 y^4+1\right) \left(2 x^2+8 x y+8 y^2+3\right) 
 \nonumber\\
 &-& 2 \left(4 x^2 y+10 x y^2+3 x+4 y^3+3 y\right) \sin{2 (x+y)}
 \nonumber\\
 &-&\left(x^2 \left(8 y^2-4\right)+4 x \left(8 y^2-1\right) y+32 y^4
 +  2 y^2+3\right)\cos{2 (x+y)} \biggr],
 \label{threeM}
\end{eqnarray}
 The spectral energy density and pressure appearing in Eq. (\ref{threeM})
depend on the two dimensionless variables $x= k\tau$ and $y = k\tau_{1}$; these expressions 
can be usefully compared with the results of Eq. (\ref{twoL}) holding in the pure de Sitter case.
The frequencies amplified in the transition from de Sitter space-time 
to the radiation epoch always obey the condition $y \ll 1$. When $x < 1$ the amplified frequencies  are still smaller than the rate of variation of the geometry: this means that to make sure that the wavelengths are larger than the Hubble radius during the radiation stage,  Eq. (\ref{threeM}) should be expanded for $ y \ll 1$ and for $x\ll1 $ with the condition $ y < x$. The leading order result of this double expansion is then given by:  
\begin{eqnarray}
\rho_{gw}^{(F)}(x,y) &=& \frac{H_{1}^4}{4 \pi^2} \frac{y^4}{x^2} \biggl[ 1 + {\mathcal O}(x^2) + {\mathcal O}\biggl(\frac{y}{x}\biggr) \biggr], 
\label{fourM}\\
p_{gw}^{(F)}(x,y) &=& - \frac{H_{1}^4}{12 \pi^2} \frac{y^4}{x^2} \biggl[ 1 + {\mathcal O}(x^2) + {\mathcal O}\biggl(\frac{y}{x}\biggr) \biggr].
\label{fiveM}
\end{eqnarray}
As expected Eqs. (\ref{fourM}) and (\ref{fiveM}) imply that the effective barotropic index is $-1/3$ while the energy density 
is positive semi-definite. The wavelengths that exited the Hubble radius during the de Sitter 
phase and reentered during the radiation epoch correspond to the limit $y \ll 1$ and $x \gg 1$; as expected,  the barotropic index goes in this case to $1/3$ even if the approach is not monotonic but oscillating as it can also be argued from Eq. (\ref{threeM}). Equations (\ref{fourM}) and (\ref{fiveM}) confirm, once more, the summary of Tab. \ref{TABLE1} in the case of ${\mathcal F}_{\mu\nu}$.

The results of Eqs. (\ref{threeM}), (\ref{fourM}) and (\ref{fiveM}) will now be compared with the analog expressions
derived from Eqs. (\ref{rhoL}) and (\ref{pL}) in the Landau-Lifshitz approach; the exact 
results in this case are:
\begin{eqnarray}
\rho_{gw}^{(L)}(x,y) &=& \frac{H_{1}^4\, y^3}{8 \pi ^2 (x+2 y)^6} \biggl[\left(2 y^4+1\right) \left(2 x^2+8 x y+8 y^2-7\right)
\nonumber\\
&+& \left(12 x y+10 y^2+7\right) \cos{2 (x+y)}
\nonumber\\
&-& 2 \left(6 x y^2-3 x+12 y^3+y\right) \sin{2 (x+y)}\biggr],
\nonumber\\
p_{gw}^{(L)}(x,y) &=& \frac{H_{1}^4\, y^3}{24 \pi ^2 (x+2 y)^6}\biggl[
\left(2 y^4+1\right) \left(2 x^2+8 x y+8 y^2-5\right)
\nonumber\\
&+& \left(12 x^2 \left(2 y^2-1\right)+4 x \left(24 y^2-7\right) y+96 y^4-18
   y^2+5\right) \cos{2 (x+y)}
 \nonumber\\
 &+& 2 \left(12 x^2 y+38 x y^2+5 x+28 y^3+5 y\right) \sin{2 (x+y)}\biggr].
\label{sixM}
\end{eqnarray}
According to Eq. (\ref{sixM}), for $x \gg 1$ and $y \gg 1$ the effective barotropic index is 
always $1/3$; this conclusion is compatible with Eq. (\ref{threeM}) in the same physical limit.
However in the limit $ y \ll 1$, $x\ll 1$ and $y< x$ the results are:
\begin{eqnarray}
\rho_{gw}^{(L)}(x,y) &=& - \frac{5 H_{1}^4}{12 \pi^2} \frac{y^4}{x^2} \biggl[ 1 + {\mathcal O}(x^2) + {\mathcal O}\biggl(\frac{y}{x}\biggr) \biggr], 
\label{sevenM}\\
p_{gw}^{(L)}(x,y) &=&  \frac{7 H_{1}^4}{12 \pi^2} \frac{y^4}{x^2} \biggl[ 1 + {\mathcal O}(x^2) + {\mathcal O}\biggl(\frac{y}{x}\biggr) \biggr].
\label{eightM}
\end{eqnarray}
According to Eqs. (\ref{sevenM}) and (\ref{eightM}) the spectral energy density is negative while 
the barotropic index is given by $-7/5$. If we consider the shifted pressure 
${\mathcal P}_{gw}^{(L)}(k,\tau)$ (see Eqs. (\ref{thirteenC}) and (\ref{sixG})) the result is different 
\begin{eqnarray}
{\mathcal P}_{gw}^{(L)}(x,y) &=&  \frac{5 H_{1}^4}{36 \pi^2} \frac{y^4}{x^2} \biggl[ 1 + {\mathcal O}(x^2) + {\mathcal O}\biggl(\frac{y}{x}\biggr) \biggr],
\label{nineM}
\end{eqnarray}
and it leads, as expected, to the more standard (i.e. $-1/3$) barotropic index. However, while 
in the case of Eqs. (\ref{fourM}) and (\ref{fiveM}) the energy density is positive and the pressure is negative 
(as it is common when the spatial gradients dominate) in the Landau-Lifshitz case the situation is reversed 
since the energy density is negative and the pressure is positive. Finally, in the case of the Brill-Hartle proposal 
the energy density is positive semi-definite and the effective barotropic index is $1/3$ when $x \gg 1$ and $y \ll 1$.  
However, in the limits $y \ll 1$, $x\ll 1$ and $y < x$ we have instead
\begin{eqnarray}
\rho_{gw}^{(L)}(x,y) &=& \frac{ H_{1}^4}{18\pi^2} \frac{y^4}{x^2} \biggl[ 1 + {\mathcal O}(x^2) + {\mathcal O}(y x) \biggr], 
\label{tenM}\\
p_{gw}^{(L)}(x,y) &=&  \frac{H_{1}^4}{6 \pi^2} \frac{y^4}{x^2} \biggl[ 1 + {\mathcal O}(x^2) + {\mathcal O}(y x) \biggr],
\label{elevenM}
\end{eqnarray}
showing that the spectral pressure is much larger than the energy density, as already discussed in the pure de Sitter case
and as it follows from the general arguments illustrated after Eq. (\ref{fourH}).
All in all the effective energy momentum tensor obtained from the second-order 
variation of the action leads to an energy density that is always gauge-invariant and positive semidefinite
exactly as argued in Tab. \ref{TABLE1}. In the Landau-Lifshitz parametrization the 
weak energy condition is violated while  the ambiguities of the Brill-Hartle approach (when applied for frequencies 
smaller than the rate of variation of the geometry) demand a completion of the energy-momentum 
pseudo-tensor in the low-frequency limit. 
\renewcommand{\theequation}{5.\arabic{equation}}
\section{Observables in the concordance scenario}
\setcounter{equation}{0}
\label{sec5}
The expectation values of the energy density [i.e. $\overline{\rho}^{(X)}_{gw}$ with $X=F,\, L,\, B$] lead to
the corresponding spectral energy densities in critical units 
\begin{equation}
\Omega_{gw}^{(X)}(k,\tau) = \frac{1}{\rho_{crit}} \, \frac{d \overline{\rho}_{gw}^{(X)}}{d \ln{k}} \equiv \frac{\rho_{gw}^{(X)}(k,\tau)}{\rho_{crit}},
\label{oneN}
\end{equation}
where $\rho_{crit} = 3 \, H^2 \, \overline{M}_{P}^2$; $\Omega_{gw}^{(X)}(k,\tau)$ together with the 
power spectra $P_{T}(k,\tau)$ and $Q_{T}(k,\tau)$ are the pivotal observables customarily employed 
in the concordance scenario to assess the energy density 
of the relic gravitons. The (less conventional) spectral pressure 
in critical units can be instead defied as:
\begin{equation}
\Sigma_{gw}^{(X)}(k,\tau) = \frac{1}{\rho_{crit}} \, \frac{d \overline{p}_{gw}^{(X)}}{d \ln{k}} \equiv \frac{p_{gw}^{(X)}(k,\tau)}{\rho_{crit}}.
\label{twoN}
\end{equation}
 $\Omega_{gw}^{(X)}(k,\tau)$ and $\Sigma_{gw}^{(X)}(k,\tau)$ will now be computed in the  different parametrizations explored so far and in the realistic situation where the evolution begins with a quasi-de Sitter phase, continues through a radiation-dominated epoch and finally arrives at a  matter-dominated stage of expansion.

 \subsection{The spectral energy density in critical units}
The properties of $\Omega_{gw}^{(F)}$ and $\Sigma_{gw}^{(F)}$ can be deduced in general terms without a specific knowledge of the evolution of the corresponding mode functions.  To illustrate this point,  Eqs. (\ref{sixF})--(\ref{sevenF}) can be inserted into  Eqs. (\ref{oneN})--(\ref{twoN}) so that the resulting expressions are:
 \begin{eqnarray}
\Omega_{gw}^{(F)}(k,\tau) &=& \frac{1}{24 \, H^2 \, a^2 }  \biggl[ Q_{T} + k^2 P_{T}\biggr], 
\label{threeN}\\
\Sigma_{gw}^{(F)}(k,\tau) &=& \frac{1}{24 \, H^2 \, a^2 }  \biggl[ Q_{T} -  \frac{k^2}{3} P_{T}\biggr].
\label{fourN}
\end{eqnarray}
To leading order in  ${\mathcal H}/k < 1$ (and even without an explicit form of the mode functions) we have that $Q_{T} = k^2\, P_{T} [ 1 + ({\mathcal H}/k)^2+ {\mathcal O}({\mathcal H}^4/k^4)]$; thus the expressions of $\Omega_{gw}^{(F)}(k,\tau)$ and $\Sigma_{gw}^{(F)}(k,\tau)$ inside the Hubble radius become:
\begin{eqnarray}
\Omega_{gw}^{(F)}(k,\tau) &=& \frac{k^2\,\, P_{T}(k,\tau)}{12 \, H^2 \, a^2 }   \biggl[ 1 + \frac{{\mathcal H}^2}{2k^2} + {\mathcal O}\biggl(\frac{{\mathcal H}^4}{k^4}\biggr) \biggr],
\label{sixN}\\
\Sigma_{gw}^{(F)}(k,\tau) &=& \frac{k^2\,\, P_{T}(k,\tau)}{36 \, H^2 \, a^2 }   \biggl[ 1 + \frac{3{\mathcal H}^2}{2k^2} + {\mathcal O}\biggl(\frac{{\mathcal H}^4}{k^4}\biggr) \biggr].
\label{sevenN}
\end{eqnarray}
The expressions of  $\Omega_{gw}^{(F)}(k,\tau)$ and $\Sigma_{gw}^{(F)}(k,\tau)$  for typical wavelengths larger than the Hubble 
radius are equally immediate since, in this limit, Eqs. (\ref{elevenF})--(\ref{twelveF}) imply 
\begin{eqnarray}
Q_{T}(k,\tau) &=& Q_{T}(k,\tau_{ex}) \biggl(\frac{a_{ex}}{a} \biggr)^4 \biggl [ 1 + {\mathcal O}\biggl(\frac{k^2}{{\mathcal H}^2}\biggr) \biggr], 
\label{eightN}\\
P_{T}(k,\tau) &=& P_{T}(k,\tau_{ex}) \biggl(\frac{a_{ex}}{a} \biggr)^4 \biggl [ 1 + {\mathcal O}\biggl(\frac{k^2}{{\mathcal H}^2}\biggr) \biggr],
\label{nineN}
\end{eqnarray}
where $P_{T}(k,\tau_{ex})$ and $Q_{T}(k,\tau_{ex})$ are the (constant) 
values of the power spectra for $k \tau_{ex} = {\mathcal O}(1)$. Inserting 
 Eqs. (\ref{eightN})--(\ref{nineN}) into Eqs. (\ref{threeN})--(\ref{fourN}) the leading-order 
 expression for $\Omega_{gw}^{(F)}(k,\tau)$ and $\Sigma_{gw}^{(F)}(k,\tau)$ are:
 \begin{eqnarray}
\Omega_{gw}^{(F)}(k,\tau) &=& \frac{k^2\,\, P_{T}(k,\tau_{ex})}{24 \, H^2 \, a^2 } \,\,\biggl[ 1 + \frac{Q_{T}(k,\tau_{ex})}{k^2\, P_{T}(k,\tau_{ex})} \biggl(\frac{a}{a_{ex}}\biggr)^4 + {\mathcal O}\biggl(\frac{k^4}{{\mathcal H}^4}\biggr) \biggr],
\label{tenN}\\
\Sigma_{gw}^{(F)}(k,\tau) &=& -\frac{k^2\,\, P_{T}(k,\tau_{ex})}{72 \, H^2 \, a^2 } \,\, \biggl[ 1 - \frac{3 Q_{T}(k,\tau_{ex})}{k^2\, P_{T}(k,\tau_{ex})} \biggl(\frac{a}{a_{ex}}\biggr)^4+ {\mathcal O}\biggl(\frac{k^4}{{\mathcal H}^4}\biggr) \biggr].
\label{elevenN}
\end{eqnarray}
If the background expands the second term inside the squared brackets at the right hand side of Eqs. (\ref{tenN})--(\ref{elevenN}) 
is always negligible and, approximately,  $\Omega_{gw}^{(F)} \simeq - \Sigma_{gw}^{(F)}/3$.
If the background contracts the second term inside the squared brackets at the right hand side of Eqs. (\ref{tenN})--(\ref{elevenN}) may become dominant and, in this case, 
$\Omega_{gw}^{(F)} \simeq \Sigma_{gw}^{(F)}$.
For wavelengths shorter that the Hubble radius the general results of Eqs. (\ref{sixN}) and (\ref{sevenN}) are, in practice, the same for all the various prescriptions and the only differences appear from the 
first correction to the leading-order result;
for illustration the next-to-leading order correction of the spectral energy density is reported in two 
relevant cases:
\begin{eqnarray}
\Omega_{gw}^{(L)}(k,\tau) &=& \frac{k^2\,\, P_{T}(k,\tau)}{12 \, H^2 \, a^2 }   \biggl[ 1 - \frac{7{\mathcal H}^2}{2k^2} + {\mathcal O}\biggl(\frac{{\mathcal H}^4}{k^4}\biggr) \biggr],
\label{twelveN}\\
\Omega_{gw}^{(B)}(k,\tau) &=& \frac{k^2\,\, P_{T}(k,\tau)}{12 \, H^2 \, a^2 }   \biggl[ 1 + \frac{{\mathcal H}^2}{k^2} + {\mathcal O}\biggl(\frac{{\mathcal H}^4}{k^4}\biggr) \biggr].
\label{thirteenN}
\end{eqnarray}
All in all, as long as we are inside the Hubble radius, 
the spectral energy density and pressure are unambiguous and do not crucially change from one 
pseudo-tensor to the others. The same is not true when the corresponding wavelengths 
are larger than the Hubble radius. 

\subsection{Explicit results in the concordance scenario}
To examine more closely the implications of the different proposals we consider a realistic evolution where a quasi-de Sitter stage of expansion ends at a time $- \tau_{r}$ and 
it is replaced by the radiation-dominated stage:
\begin{equation}
a_{r}(\tau) = \frac{\beta \tau + (\beta+ 1) \tau_{r}}{\tau_{r}}, \qquad x(\tau) = k \biggl[ \tau + \frac{\beta +1}{\beta} \tau_{r} \biggr],
\label{oneO}
\end{equation}
where $\beta = (1 -\epsilon)^{-1}$ is a numerical factor required for the continuity of the scale factors in the quasi-de Sitter
stage and $\epsilon = - \dot{H}/H^2$ is the conventional slow-roll parameter. The inflationary phase ends for $\tau= - \tau_{r}$ and the scale factor is normalized as $a_{r}(- \tau_{r}) =1$. The evolution dictated by Eq. (\ref{oneO}) 
lasts until $\tau_{m}$ when the matter dominated stage begins:
\begin{equation}
a_{m}(\tau) = \frac{[\beta (\tau + \tau_{m}) + 2 (\beta+ 1) \tau_{r}]^2}{4 \tau_{r}[ \beta \tau_{m} + (\beta+1) \tau_{r}]}, \qquad y(\tau) = k \biggl[ \tau + \tau_{m} + 2 \frac{\beta +1}{\beta} \tau_{r} \biggr],
\label{threeO}
\end{equation}
where $a_{m}(\tau_{m}) = a_{r}(\tau_{m})$ and $a_{m}^{\prime}(\tau_{m}) = a_{r}^{\prime}(\tau_{m})$.
From Eqs. (\ref{oneO}) and (\ref{threeO})  the power 
spectra can be derived before (i.e. $\tau < \tau_{m}$)  and after (i.e.  $\tau>\tau_{m}$) the dominance of matter:
\begin{eqnarray}
P^{(r)}_{T}(k,\tau, \tau_{r}) &=& \overline{P}_{T}(k,\tau_{r}) \frac{\sin^2{x(\tau)}}{|x(\tau)|^2}, \qquad \tau< \tau_{m},
\label{fourO}\\
P^{(m)}_{T}(k, \tau, \tau_{r}, \tau_{m}) &=& 9 \overline{P}_{T}(k,\tau_{r})\biggl[ \frac{\cos{y(\tau)}}{y^2(\tau)} - \frac{\sin{y(\tau)}}{y^3(\tau)}\biggr]^2, \qquad \tau > \tau_{m}.
\label{fiveO}
\end{eqnarray}
When the relevant wavelengths exceed the Hubble radius the general expressions of 
Eqs. (\ref{fourO}) and (\ref{fiveO}) coincide i.e. 
\begin{equation}
\lim_{|k\tau| \ll 1}  P^{(r)}_{T}(k,\tau, \tau_{r}) = \lim_{|k\tau| \ll 1} P^{(m)}_{T}(k, \tau, \tau_{r}, \tau_{m}) = \overline{P}_{T}(k,\tau_{r}),
\label{fiveOa}
\end{equation}
where $\overline{P}_{T}(k,\tau_{r})$ denotes the (constant) inflationary power spectrum:
\begin{equation}
\overline{P}_{T}(k,\tau_{r}) = 2^{ 2 \nu} \frac{\Gamma^2(\nu)}{\pi^3} \biggl(\frac{H_{r}}{\overline{M}_{P}}\biggr)^2 
\, |k\tau_{r}|^{ 3  - 2 \nu}, \qquad \nu = \frac{(3 - \epsilon)}{ 2( 1- \epsilon)}.
\label{twoO}
\end{equation}
Inserting Eqs. (\ref{fourO})--(\ref{fiveO}) into the general expressions of Eqs. (\ref{threeN})--(\ref{fourN}) 
 the spectral energy density and pressure inside and beyond the Hubble radius can be obtained and they are:
\begin{eqnarray}
\Omega_{gw}^{(F)}(k,\tau) &=& \frac{\overline{P}_{T}(k,\tau_{r})}{24},\qquad  \Sigma_{gw}^{(F)}(k,\tau) = \frac{\overline{P}_{T}(k,\tau_{r})}{72},\qquad |k\tau| \gg 1, 
 \label{fiveOb}\\
 \Omega_{gw}^{(F)}(k,\tau) &=& \frac{\overline{P}_{T}(k,\tau_{r})}{24} |k\tau|^2 ,\qquad  \Sigma_{gw}^{(F)}(k,\tau) = - \frac{\overline{P}_{T}(k,\tau_{r})}{72} |k\tau|^2,\qquad |k\tau| \ll 1.
 \label{fiveOc}
 \end{eqnarray}
 Equations (\ref{fiveOb})--(\ref{fiveOc}) arise as limits of concrete expressions (holding for a specific 
 form of the mode functions) and they agree with the results of Eqs. (\ref{sixN})--(\ref{sevenN}) and (\ref{tenN})--(\ref{elevenN}) that are instead derived as approximated expressions\footnote{It follows from Eq. (\ref{fourO}), $P_{T}(k,\tau) \to \overline{P}_{T}(k, \tau_{r})$ for $|k\tau| \ll 1$ while $P_{T}(k,\tau) \to \overline{P}_{T}(k, \tau_{r})/2$ for $|k\tau| \gg 1$ since, in this limit, $ \sin^2{x(\tau)} \to 1/2$. } of the general results (\ref{threeN})--(\ref{fourN}). A similar discussion can be repeated in the matter-dominated stage of expansion (i.e. $\tau > \tau_{m}$) where the limits of the concrete expressions read:
\begin{eqnarray}
\Omega_{gw}^{(F)}(k,\tau) &=& \frac{3}{32\, |k\tau|^2}\overline{P}_{T}(k,\tau_{r}),\qquad  \Sigma_{gw}^{(F)}(k,\tau) = \frac{\overline{P}_{T}(k,\tau_{r})}{32 \, |k\tau|^2},\qquad |k\tau| \gg 1, 
 \label{fiveOd}\\
 \Omega_{gw}^{(F)}(k,\tau) &=& \frac{\overline{P}_{T}(k,\tau_{r})}{96} |k\tau|^2 ,\qquad  \Sigma_{gw}^{(F)}(k,\tau) = - \frac{\overline{P}_{T}(k,\tau_{r})}{288} |k\tau|^2,\qquad |k\tau| \ll 1.
 \label{fiveOe}
 \end{eqnarray}
The results of Eqs.  (\ref{fiveOb})--(\ref{fiveOc}) and (\ref{fiveOd})--(\ref{fiveOe}) can be compared with 
the analog results obtainable in the Landau-Lifshitz parametrization. Consider first the radiation 
phase  (i.e. $\tau < \tau_{m}$) where 
\begin{equation}
\Omega_{gw}^{(L)}(k,\tau) 
= - \frac{5}{72} |k\, \tau|^2 \overline{P}_{T}(k,\tau_{r}),\qquad 
\Sigma_{gw}^{(L)}(k,\tau) =  \frac{5}{216}   |k\, \tau|^2  \overline{P}_{T}(k,\tau_{r}),\qquad  |k\, \tau| \ll 1.
\label{sevenO}
\end{equation}
Equation (\ref{sevenO}) gives the spectral energy density and the spectral pressure in critical units 
during the radiation epoch (i.e. 
for $\tau_{m} > \tau > - \tau_{r}$) and when the relevant wavelengths are larger than the Hubble radius (i.e.  
$|k \tau| \ll 1$ with $|k \tau_{r}| \ll 1$). Similarly we can also deduce the spectral energy density 
during the matter stage: 
\begin{equation}
\Omega_{gw}^{(L)}(k,\tau) = - \frac{11}{480} \,|k\tau|^2 \,\overline{P}_{T}(k,\tau_{r}),\qquad 
\Sigma_{gw}^{(L)}(k,\tau) =  \frac{11}{1440} \, |k\tau|^2  \overline{P}_{T}(k,\tau_{r}), \qquad |k\, \tau| \ll 1.
\label{nineO}
\end{equation}
Equations (\ref{sevenO}) and (\ref{nineO})  imply that the spectral 
energy density in critical units is negative even if it is still true that $\Sigma_{gw}^{(L)}(k,\tau) = - \Omega_{gw}^{(L)}(k,\tau)/3$.
If we compare Eqs. (\ref{fiveOc}) and (\ref{fiveOe}) with Eqs. (\ref{sevenO}) and (\ref{nineO}) we see that 
the overall sign of the energy density and of the pressure are completely reversed. While Eqs. (\ref{fiveOc}) and (\ref{fiveOe}) could be obtained on a general ground without specifying the details of the mode functions, the overall normalization of Eqs. (\ref{sevenO}) and (\ref{nineO}) does depend on 
the details of the mode function and not only on the 
relation between the spectral energy density and the power spectrum.

\subsection{Stationary random process inside the Hubble radius}
If the tensor amplitude is not a quantum field operator but it describes a stationary random process 
the spatial variation can be approximately neglected by 
focussing on the conformal time dependence:
\begin{equation}
h_{ij}(\tau) = \sum_{\lambda} e_{ij}^{(\lambda)} \, h_{\lambda}(\tau), \qquad 
h_{\lambda}(\tau) = \frac{1}{\sqrt{2 \pi}} \int_{- \infty}^{\infty} e^{i \,\omega\,\tau}\, h_{\lambda}(\omega)\, d\tau.
\label{elevenO}
\end{equation}
In this case  the spectral energy density  is only determined 
by the temporal variation of the tensor amplitude and can be deduced within 
the Brill-Hartle scheme with the caveat that the result 
will only apply inside the Hubble radius. 
If the random process is stationary, 
by definition the autocorrelation function will only depend on the time difference 
at which the two amplitudes are evaluated, i.e. $\langle h_{\lambda}(\tau) h_{\lambda^{\prime}}(\tau)\rangle = \delta_{\lambda\, \lambda^{\prime}} \Gamma(\tau - \tau^{\prime})$.  The temporal autocorrelation implies that in Fourier space 
\begin{equation}
\langle h_{\lambda}(\omega) h_{\lambda^{\prime}}(\omega^{\prime})\rangle = S_{h}(\omega) \delta_{\lambda\, \lambda^{\prime}} \delta(\omega+  \omega^{\prime}),
\label{twelveO}
\end{equation}
where $S_{h}(\omega)$ is the spectral density. Using Eqs.(\ref{elevenO}) inside  Eq. (\ref{rhoB}) 
the expectation value of the energy density in the Brill-Hartle-Isaacson approach follows
from the stochastic average of Eq. (\ref{twelveO}):
\begin{equation}
\overline{\rho}_{gw} = \frac{1}{4 \ell_{P}^2 a^2 } \langle \, \partial_{\tau} \, h_{ij}\, \partial_{\tau} \, h^{i\,j} \rangle = \frac{1}{2\, \pi a^2 \ell_{P}^2} \int \frac{d\, k}{k} \, k^3\, S_{h}(k).
\label{thirteenO}
\end{equation}
Since the expression of $\Omega_{gw}(k,\tau)$  inside the Hubble radius is unambiguous, 
Eq. (\ref{thirteenO}) implies then a specific relation between 
the power spectrum  $P_{T}$, the spectral amplitude $S_{h}$ and the spectral 
energy density in critical units:
\begin{equation}
\Omega_{gw}(k) = \frac{k^2}{ 12\, H^2\, a^2} P_{T}(k) \equiv \frac{k^{3}}{6 \pi H^2 a^2 } \, S_{h}(k).
\label{fourteenO}
\end{equation}
If we pass from the angular frequencies $\omega$ to the frequencies $\nu$ (and recall that in the natural units 
adopted here $\omega = k = 2 \pi \nu$) Eq. (\ref{fourteenO}) can also be phrased as:
\begin{equation}
P_{T}(\nu) = 4 \nu S_{h}(\nu) , \qquad \Omega_{gw}(\nu) =  \frac{4 \pi^{2}\,\nu^3}{3 H^2 a^2 } \, S_{h}(\nu).
\label{fifteenO}
\end{equation}
The result of Eq. (\ref{fifteenO}) demonstrates that the tensor amplitudes can be considered as isotropic random fields characterized by stationary autocorrelation functions. In this case Eqs. (\ref{sixE}), (\ref{sevenE}) and (\ref{twelveO}) 
must be viewed as averages of classical stochastic processes not necessarily related to quantum field
operators. 

\subsection{Frame-invariance of the effective action}
The effective energy-momentum pseudo-tensor shall now be  evaluated 
in a generalized Jordan frame where the scalar-tensor action reads: 
\begin{equation}
S_{J} = \int d^4 x\, \sqrt{ - G}\,\biggl[ -  \frac{A(\varphi)}{ 2 \ell_{P}^2}\, R_{J} + \frac{B(\varphi)}{2} G^{\alpha\beta} \partial_{\alpha} \varphi \partial_{\beta} \varphi 
- V(\varphi) \biggr],
\label{oneP}
\end{equation}
where $A(\varphi)$ and $B(\varphi)$ are dimensionless and depend on the scalar field $\varphi$. The second-order variation of Eq. (\ref{oneP}) 
 can be easily obtained by repeating the same steps leading to Eqs. (\ref{twoB}) and (\ref{twoBa})--(\ref{twoBb}) and the 
 result is:
\begin{eqnarray}
 \delta^{(2)}_{t} S_{J} &=&\int d^{4} x  \, \biggl\{ \frac{1}{2 \ell_{P}^2}\,\biggl[ A(\varphi) \, \overline{G}^{\alpha\beta} \, \overline{{\mathcal Z}}_{\alpha\beta} \,\,\delta^{(2)}_{t} \sqrt{-G} + A(\varphi)\, \sqrt{ -\overline{G}} \biggl( \delta^{(2)}_{t} G^{\alpha\beta} \,\overline{{\mathcal Z}}_{\alpha\beta} 
 \nonumber\\
 &+&
 \delta^{(1)}_{t} G^{\alpha\beta} \,  \delta^{(1)}_{t} {\mathcal Z}_{\alpha\beta} + \overline{G}^{\alpha\beta}\,\,\delta^{(2)}_{t} {\mathcal Z}_{\alpha\beta} \biggr)
\nonumber\\
&-& \delta^{(2)}_{t}\biggl( \sqrt{-G} \, G^{\alpha\beta} \, \Gamma_{\alpha\lambda}^{\,\,\,\,\,\lambda} \, \partial_{\beta} A\biggr) 
+  \delta^{(2)}_{t}\biggl( \sqrt{-G} \, G^{\alpha\beta} \, \Gamma_{\alpha\beta}^{\,\,\,\,\,\lambda} \, \partial_{\lambda} A\biggr) \biggr]
\nonumber\\
&+&  \delta_{t}^{(2)} \sqrt{-G} \biggl(\frac{B}{2} \overline{G}^{\alpha\beta} \partial_{\alpha} \varphi \partial_{\beta} \varphi 
- V(\varphi)\biggr) + \sqrt{-\overline{G}} \frac{B}{2} \delta_{t}^{(2)} G^{\alpha\beta} \,\partial_{\alpha} \varphi \partial_{\beta}\varphi \biggr\}.
\label{onePa}
\end{eqnarray}
Equation (\ref{onePa}) contains comparatively more terms than the analog results
valid in the case $A\to 1$ (see e.g.  (\ref{twoB})) where various contributions disappear and are replaced by a pair of total derivatives 
that do not affect the final result. After some lengthy but straightforward algebra the explicit form of the second-order action reads:
\begin{eqnarray}
S_{t\, J} &=& \delta^{(2)} S_{J} = \frac{1}{8 \ell_{P}^2} \int d^{4}x \sqrt{-\overline{G}} \, \overline{G}^{\alpha\beta} \, \, A(\varphi) \, \partial_{\alpha} h^{\,\,\,\,(J)}_{i\,j}
\partial_{\beta} h^{(J)\,\,\,i\,j},
\nonumber\\
&-& \frac{1}{8\ell_{P}^2} \int d^{4} x \, a^2_{J} A(\varphi) \,h^{\,\,\,\,(J)}_{k\ell} \,\, h^{(J)\,\,k \ell} \bigg[ 4 {\mathcal H}^{\prime} + 2 {\mathcal M}^{\prime} 
+ 2 ({\mathcal H}^2 + {\mathcal H} {\mathcal M} + {\mathcal M}^2) 
\nonumber\\
&+& \frac{ 2 \ell_{P}^2}{A} \biggl( \frac{B}{2} \varphi^{\prime \, \,2} - V\, a_{J}^2\biggr)\biggr],
\label{onePb}
\end{eqnarray}
where ${\mathcal M} = A^{\prime}/A$.
The tensor amplitude $h^{(J)}_{ij}$ entering Eq. (\ref{onePb}) is defined directly
 in the Jordan frame, i.e.  $\delta_{t}^{(1)} G_{ij} = -a_{J}^2 \, h^{\,(J)}_{ij}$;  $a_{J}$ is the scale factor appearing in 
 the $J$-frame, i.e. $\overline{G}_{\alpha\beta} = a^2_{J} \,\,\eta_{\alpha\beta}$.
The expression inside the squared bracket of Eq. (\ref{onePb}) vanishes identically since it 
corresponds to the $(ij)$ component of the  background equations derived from the extremization of the action (\ref{onePa}) with 
respect to the variation of the metric.  By considering the tensor amplitude $h_{ij}^{\,\,(J)}$ and the background metric as independent variables the effective energy-momentum tensor in the $J$-frame follows from Eq. (\ref{onePb}) and it is:
\begin{equation}
T_{\mu\nu}^{(J)} = \frac{A}{4 \ell_{P}^2} \biggl[ \partial_{\mu} h^{\,\,\,\,(J)}_{k\ell} \partial_{\nu} \overline{h}^{(J)\,\,\,\, k\ell} - 
\frac{1}{2} \overline{G}_{\mu\nu} \biggl( \overline{G}^{\alpha\beta} \partial_{\alpha} h^{\,\,\,\,(J)}_{k\ell} \partial_{\beta} \overline{h}^{(J)\,\,\,\, k\ell}\biggr) \biggr],
\label{threeP}
\end{equation}
in full analogy with the result of Eq. (\ref{fourB}). The energy density in the $J$-frame becomes
\begin{equation}
\rho_{gw}^{(J)} = \frac{A}{8 \ell_{P}^2 a^2_{J}} \biggl[ \partial_{\tau} h^{\,\,\,\,(J)}_{k\ell} \partial_{\tau} \overline{h}^{(J)\,\,k\ell}
+ \partial_{m} h^{\,\,(J)}_{k\ell}\partial^{m} \overline{h}^{(J)\,\,k\ell} \biggr].
\label{fourP}
\end{equation}
The conformal rescaling $A \, G_{\alpha\beta} = g_{\alpha\beta}$ brings the action (\ref{onePb}) from the $J$-frame  the Einstein frame:
\begin{equation}
a_{J}^2 \,A = a^2, \qquad A \,a^2_{J} h^{(J)}_{ij} = a^2\, h_{ij},
\label{fiveP}
\end{equation}
where the first equality follows from the conformal rescaling of the 
background (i.e. $A \overline{G}_{\alpha\beta} = \overline{g}_{\alpha\beta}$) 
while the second equality is implied by the relation between the first-order tensor fluctuations in the two frames 
(i.e. $A \delta_{t}^{(1)} G_{ij} = \delta_{t}^{(1)} g_{ij}$). Equation. (\ref{fiveP}) also requires that $h_{ij} = h^{\,(J)}_{ij}$ so that the action of Eq. (\ref{twoB}) coincides with the Einstein frame action of Eq. (\ref{threeB}). As a consequence, the energy densities in the two frames are related as:
\begin{equation}
\rho_{gw}^{(J)} = A^2 \, \rho_{gw}^{(E)} \equiv\, \frac{\sqrt{- \overline{g}}}{\sqrt{ - \overline{G}}} \rho_{gw}^{(E)},
\label{sixP}
\end{equation}
where $\rho_{gw}^{(E)}$ coincides with Eq. (\ref{rhoF}). Since the energy density of a radiation 
plasma also scales as $\rho_{r}^{(J)} = A^2 \, \rho^{(E)}_{r}$, Eq. (\ref{fiveP}) implies that $\rho_{gw}^{(J)}/\rho_{r}^{(J)}= \rho_{gw}^{(E)}/\rho_{r}^{(E)}$. This 
observation ultimately implies that the spectral energy density in critical units is  the same in the two conformally 
related frames (i.e.  $\Omega_{gw}^{(J)} = \Omega_{gw}^{(E)}$).  Let us remark, as we close, that the class of scalar-tensor theories of Eq. (\ref{oneP}) is purely illustrative and the
effective action of the relic gravitons may also inherit further parity-violating contribution \cite{twentysix,twentyseven}; in this case a more general form 
of the effective action has been proposed in \cite{twentyeight} and it is relevant for the 
description of the polarized backgrounds of relic gravitons. This development is however 
not central to the present considerations.
\renewcommand{\theequation}{6.\arabic{equation}}
\section{Concluding remarks}
\setcounter{equation}{0}
\label{sec6}
The energy density of the relic gravitons is not univocally and unambiguously defined. The various suggestions proposed so far 
coincide when the rate of variation of the background geometry is smaller than the frequency 
of the corresponding gravitons. However, in cosmological backgrounds the rate of variation 
of the space-time curvature can also exceed the typical frequencies 
of the gravitons. The energy-momentum pseudo-tensor of the relic gravitons should fulfil 
four plausible criteria: it should be frame-invariant 
and gauge-invariant, it should not violate the weak energy condition and it should be derived in general terms, i.e. 
without explicitly demanding that the rate of variation of the background 
geometry is either faster or slower than the frequencies of the corresponding gravitons.
An energy-momentum pseudo-tensor with these features
follows from the effective action of the relic gravitons 
by considering the tensor fluctuations and the background metric as independent 
variables. In its simplest realization the effective action coincides with 
the result of Ford-Parker and it is defined in all the relevant physical regimes.  
The spectral energy density in critical units derived within this approach is gauge-invariant and 
 frame invariant since its value in two conformally related frame does not change.
  
If we assume, a priori, that the typical frequencies of the gravitons must exceed the rate of 
variation of the geometry we are implicitly following the logic of the Brill-Hartle-Isaacson pseudo-tensor
whose results are applicable when the wavelengths of the corresponding 
gravitons are shorter than the Hubble radius. This proposal can be extended to encompass 
wavelengths larger than the Hubble radius; if this is done the Brill-Hartle-Isaacson result coincides
 with the effective energy-momentum tensor derived from the second-order variation of the 
action.  Finally the Landau-Lifshitz pseudo-tensor does not assume that the frequencies must exceed 
the rate of variation of the geometry but its  expression depends explicitly on the expansion rates.
Hence the actual results for the energy density and of the pressure easily
follow from the specific evolution of the mode functions but they are difficult to 
assess in general terms. In various realistic and semi-realistic situations the energy density 
computed in the Landau-Lifshitz approach always becomes negative when the typical wavelengths 
are larger than the Hubble radius. It seems difficult to attribute a profound physical 
significance to this occurrence: since we explicitly demonstrated that there exist effective pseudo-tensors
not leading to a negative energy density, there are no reasons 
to conclude that relic gravitons must inevitably violate the weak energy condition
as they evolve beyond the Hubble radius.

All in all the effective action of the relic gravitons discussed here 
leads to a computable energy-momentum pseudo-tensor 
that can be assessed in the asymptotic physical regimes even without a 
detailed knowledge of the background evolution. In this context the energy density is positive 
semi-definite and the whole description can be easily extended to a conformally related frame. 
The other strategies examined in this investigation give reasonable results
only when the relevant wavelengths are shorter than the 
Hubble radius. Even if the present conclusions have been reached in the framework 
of a quantum mechanical averaging scheme rooted in the properties of the relic gravitons,
we argued that the same conclusions can be obtained by considering the tensor amplitudes 
as isotropic random fields characterized by stationary autocorrelation functions.

\section*{Acknowledgements}

The author wishes to thank T. Basaglia, A. Gentil-Beccot and S. Rohr of the CERN Scientific Information Service 
for their kind assistance.

\end{document}